\documentclass[sigconf]{acmart}

\usepackage[linesnumbered,ruled,vlined]{algorithm2e}
\usepackage{amsthm}
\usepackage{multirow}
\usepackage{enumitem}

\definecolor{ColorRoundE}{rgb}{0.85,0.03,0.19}
\definecolor{ColorRoundD}{rgb}{0.376,0.619,0.843}
\definecolor{ColorRoundF}{rgb}{0.54,0.17,0.87}

\newcommand{\CHENG}{}
\newcommand{\cheng}{}
\newcommand{\chengr}{}
\newcommand{\chengrr}{}
\newcommand{\chengB}{}

\newcommand{\revision}{}

\def\CR{1}

\newtheorem{problem}{Problem}
\newtheorem{subproblem}{Sub-Problem}

\newtheorem{theorem}{Theorem}
\newtheorem{definition}{Definition}
\newtheorem{lemma}{Lemma}

\newtheorem{property}{Property}

\AtBeginDocument{%
  \providecommand\BibTeX{{%
    \normalfont B\kern-0.5em{\scshape i\kern-0.25em b}\kern-0.8em\TeX}}}

\copyrightyear{2024}
\acmYear{2024}
\setcopyright{acmlicensed}\acmConference[SIGMOD '24]{Proceedings of the 2024 International Conference on Management of Data}{June 11--16, 2024}{Santiago, Chile}
\acmBooktitle{Proceedings of the 2024 International Conference on Management of Data (SIGMOD '24), June 11--16, 2024, Santiago, Chile}
\acmDOI{}

\acmPrice{15.00}
\acmISBN{978-1-4503-XXXX-X/18/06}

\begin{document}
\sloppy


\title{Fast Maximal Quasi-clique Enumeration: A Pruning and Branching Co-Design Approach}

\author{Kaiqiang Yu}
\affiliation{%
  \institution{Nanyang Technological University}
  \country{Singapore}
}
\email{kaiqiang002@e.ntu.edu.sg}
\author{Cheng Long}
\authornote{Cheng Long is the corresponding author.}
\affiliation{%
  \institution{Nanyang Technological University}
  \country{Singapore}
}
\email{c.long@ntu.edu.sg}

\begin{abstract}
    
    Mining cohesive subgraphs from a graph is a fundamental problem in graph data analysis. One notable cohesive structure is $\gamma$-quasi-clique (QC), where each vertex connects at least a fraction $\gamma$ of the other vertices inside. Enumerating maximal $\gamma$-quasi-cliques (MQCs) of a graph has been widely studied and used for many applications such as community detection and significant biomolecule structure discovery. One common practice of finding all MQCs is to (1) find a set of QCs containing all MQCs and then (2) filter out non-maximal QCs. While quite a few algorithms have been developed (which are branch-and-bound algorithms) for finding a set of QCs that contains all MQCs, all focus on sharpening the pruning techniques and devote little effort to improving the branching part. As a result, they provide no guarantee on pruning branches and all have the worst-case time complexity of $O^*(2^n)$, where $O^*$ suppresses the polynomials and $n$ is the number of vertices in the graph. In this paper, we focus on the problem of finding a set of QCs containing all MQCs but deviate from further sharpening the pruning techniques as existing methods do. We pay attention to both the pruning and branching parts and develop new pruning techniques and branching methods that would suit each other better towards pruning more branches both theoretically and practically. Specifically, we develop a new branch-and-bound algorithm called \texttt{FastQC} based on newly developed pruning techniques and branching methods, which improves the worst-case time complexity to $O^*(\alpha_k^n)$, where $\alpha_k$ is a positive real number strictly \emph{smaller} than 2. Furthermore, we develop a divide-and-conquer strategy for boosting the performance of \texttt{FastQC}. Finally, we conduct extensive experiments on both real and synthetic datasets, and the results show that our algorithms are up to two orders of magnitude faster than the state-of-the-art on real datasets. 
\end{abstract}

\ccsdesc[100]{Mathematics of computing~Graph algorithms}

\keywords{cohesive subgraph enumeration; quasi-clique; branch-and-bound}

\maketitle

\section{Introduction}

Cohesive subgraph mining is a fundamental problem in graph data analysis. For a given graph, it aims to find \emph{dense/cohesive} subgraphs that carry interesting information for solving practical problems~\cite{fang2021cohesive}. One notable cohesive structure is $\gamma$-quasi-clique (QC)~\cite{pei2005mining,jiang2009mining,zeng2006coherent,liu2008effective,khalil2022parallel,guo2020scalable,sanei2018enumerating}, which is a natural generalization of clique~\cite{bron1973algorithm,tomita2006worst,eppstein2011listing,conte2020sublinear,li2022one,chang2019efficient}. Specifically, QC requires that each vertex connects at least a fraction $\gamma$ of the other vertices inside, where $\gamma$ is a fraction between 0 and 1. 
One of the fundamental QC mining problems, {\chengr which we call MQCE,} is to enumerate all large Maximal Quasi-Cliques (MQCs) with the number of vertices inside at least a threshold $\theta$ for a given graph~\cite{pei2005mining,jiang2009mining,zeng2006coherent,liu2008effective,khalil2022parallel,guo2020scalable}.

{\chengr The MQCE problem}
has been widely studied in the past~\cite{pei2005mining,jiang2009mining,zeng2006coherent,liu2008effective,khalil2022parallel,guo2020scalable} and used for various applications such as discovering biologically relevant functional groups~\cite{harley2001uniform,bader2003automated,bhattacharyya2009mining,bu2003topological}, finding social communities~\cite{fang2020survey,guo2022maximal}, detecting anomaly~\cite{yu2021graph,tanner2010koobface,weiss2015tracking}, etc. 
{\cheng For example, authors in~\cite{pei2005mining} {\cheng conduct a case study that finds} biologically relevant functional groups by mining large {\cheng MQCs which have} the size at least a threshold {\cheng and appear in each graph from} a set of protein-protein interaction and gene-gene interaction graphs.
The rationale is that for a functional group of proteins, each of them interacts with most of the rest, which would form a QC {\cheng likely}~\cite{pei2005mining}. 
{\cheng Another example is that the authors in~\cite{guo2022maximal} conduct a case study that finds meaningful communities by mining large MQCs from graphs built on publication data.}

}

{\cheng 
\smallskip
\noindent\textbf{Challenges and Existing Methods.}
{\chengr The MQCE problem}
is challenging, which is evidenced by several facts. First, this problem is NP-hard~\cite{pastukhov2018maximum}. Second, the problem of checking whether a QC is a maximal one is also NP-hard~\cite{sanei2018enumerating}. Third, QCs do not satisfy the \emph{hereditary property} (since a subgraph of a QC is not always a QC). As a result, many advanced techniques that have been developed for enumerating subgraphs that satisfy the hereditary property (e.g., $k$-plexes, $s$-defective cliques, etc.) cannot be utilized for this problem~\cite{zhou2020enumerating,yu2022maximum,chen2021computing}.  
{\revision {\chengr One common practice of solving the MQCE problem involves two steps: (1) it finds a set of QCs that contains all MQCs, which may involve non-maximal QCs; (2) it filters out non-maximal QCs from those QCs found in the first step~\cite{liu2008effective,guo2020scalable,guo2022maximal}. 
This is mainly because checking whether a QC is maximal directly is NP-hard~\cite{sanei2018enumerating}.
Therefore, we decompose the MQCE problem into two sub-problems, namely MQCE-S1 and MQCE-S2, each for a step involved in solving the MQCE problem.
Existing studies~\cite{liu2008effective,guo2020scalable,guo2022maximal} usually focus on the MQCE-S1 problem since the MQCE-S2 problem can be solved efficiently with existing techniques for the \emph{set containment query}, for which there exists a rich literature~\cite{rivest1976partial,charikar2002new,bevc2009using,savnik2013index,savnik2021data}.
}
%

Quite a few algorithms have been developed for the 
{\chengr MQCE-S1 problem,}
including \texttt{Crochet}~\cite{pei2005mining,jiang2009mining}, \texttt{Cocain}~\cite{zeng2006coherent}, \texttt{Quick}~\cite{liu2008effective} and \texttt{Quick+}~\cite{khalil2022parallel,guo2020scalable}.}
They all correspond to \emph{branch-and-bound} (BB) algorithms. Specifically, they recursively partition the search space (i.e., the set of all possible vertex sets) to multiple sub-spaces with a \emph{branching} method - each sub-space corresponds to a \emph{branch} and develop techniques for pruning some branches that hold no MQCs. These algorithms share the branching method, which is the one behind a classic \emph{set-enumeration} (SE) tree and thus we call it the \emph{SE branching} method. They differ in their pruning techniques.
One insufficiency that is suffered by all existing methods~\cite{pei2005mining,jiang2009mining,zeng2006coherent,liu2008effective,khalil2022parallel,guo2020scalable} is that they devote little effort to improving the branching part, i.e., they uniformly adopt the SE branching method, and focus \emph{solely} on sharpening the pruning techniques. As a result, the pruning part and the branching part are often not well optimized \emph{jointly} towards the goal of pruning as many branches as possible. In fact, none of these methods can provide theoretical guarantee on pruning branches. This is reflected by the fact all of them have the worst-case time complexity of $O^*(2^n)$, where $O^*$ suppresses the polynomials and $n$ denotes the number of vertices of the graph.

\smallskip
\noindent\textbf{New Methods.}
{\revision In this paper, we 
{\chengr focus on the MQCE-S1 problem}
but deviate from the direction of sharpening pruning techniques further while adopting the SE branching method as existing studies all pursue~\cite{pei2005mining,jiang2009mining,zeng2006coherent,liu2008effective,khalil2022parallel,guo2020scalable}.} We aim to develop new pruning techniques and branching methods that would suit each other better towards pruning more branches \emph{both theoretically and practically}. Specifically, we first develop a pruning technique, which is based on a necessary condition for a branch to hold QCs (i.e., if a branch does not satisfy the condition, we can prune the branch). One nice property of the pruning technique is that if a branch with a \emph{partial set} $S$ can be pruned, then all other branches with the partial sets as \emph{supersets} of $S$ can also be pruned. Here, a partial set of a branch means the set of vertices that are included in all vertex sets under this branch. To fully unleash the power of this new pruning technique, we adopt a branching method that is \emph{symmetric} to the SE branching method. We call this new branching method \emph{Sym-SE branching}. Given a current branch, Sym-SE branching would create a series of branches such that the following branches have their partial sets as \emph{supersets} of those of the preceding branches. Therefore, once we find that a branch can be pruned by our new pruning technique, all branches that follow this branch in the series can be pruned as well. We further observe that SE branching and Sym-SE branching can be jointly applied in certain cases so that more branches can be pruned. We call the resulting branching the \emph{Hybrid-SE branching} method. We show that a BB algorithm that is based on our newly developed pruning technique and branching methods, which we call \texttt{FastQC}, would have a worst-case time complexity of $O(n\cdot d \cdot \alpha_k^n)$ (i.e., $O^*(\alpha_k^n)$) where $d$ is the maximum degree of a vertex and $\alpha_k$ is strictly smaller than 2 and depends on the value of $k$, e.g., $\alpha_k=1.769$ when $k=2$. 
%
}

In addition, we adapt a \emph{divide-and-conquer (DC)} strategy for boosting the efficiency and scalability of \texttt{FastQC}.
Basically, it divides the whole graph into multiple smaller ones and then runs \texttt{FastQC} on each of them. Furthermore, we develop some new pruning techniques to shrink the constructed smaller graphs for better efficiency. In summary, the resulting algorithm called \texttt{DCFastQC} would invoke \texttt{FastQC} multiple times, each on a smaller graph (compared with the original graph), and thus the scalability is improved.
We note that this DC strategy has been widely used for enumerating subgraphs~\cite{guo2020scalable, yu2022maximum, zhou2020enumerating, khalil2022parallel}. Our technique differs from existing ones in (1) the way of how a graph is divided~\cite{guo2020scalable,khalil2022parallel}; and/or (2) the techniques for shrinking the smaller graphs~\cite{guo2020scalable, yu2022maximum, zhou2020enumerating, khalil2022parallel}.

\smallskip
\noindent\textbf{Contributions.} Our contributions are summarized below.
\begin{itemize}[leftmargin=*]
    \item We propose a new BB algorithm called \texttt{FastQC} {\revision for {\chengr the MQCE-S1 problem, i.e., the problem of} finding a set of QCs containing all MQCs}, which is based on our newly developed pruning technique and branching methods. \texttt{FastQC} 
    {\cheng has the} worst-case time complexity of $O(n\cdot d\cdot \alpha_k^n)$ with $\alpha_k<2$,
    {\cheng which breaks the long-standing bottleneck time complexity of $O^*(2^n)$~\footnote{{\cheng We note that there has been some recent progress of improving the time complexity for enumerating some subgraphs that satisfy the hereditary property (e.g., $k$-plex and $s$-defective clique)~\cite{zhou2020enumerating,chen2021computing}, but they cannot be used for our problem since QCs do not satisfy the property. To our best knowledge, this is the first breakthrough of worst-case time complexity of enumerating subgraphs that do not satisfy the hereditary property.}}.
    }
    %
    (Section~\ref{sec:FastQC})
    
    \item We further introduce a divide-and-conquer strategy, called \texttt{DC}, for boosting the performance of \texttt{FastQC}. When applying \texttt{DC} to \texttt{FastQC}, the worst-case time complexity becomes $O(n\cdot \omega d^2\cdot \alpha_k^{\omega d})$ where $\omega$ is the degeneracy of the given graph. This is better than that of \texttt{FastQC} on certain real-world graphs (e.g., those sparse graphs with $\omega<<n$ or $d<<n$). 
    (Section~\ref{sec:DC_Framework})
    
    \item We conduct extensive experiments on both real and synthetic datasets to evaluate the efficiency and scalability {\cheng of our algorithms}, e.g., \texttt{DCFastQC} is up to {\cheng two} orders of magnitude faster than the state-of-the-art \texttt{Quick+} {\cheng on real datasets}. 
    (Section~\ref{sec:exp})
\end{itemize}

For the rest of the paper, we {\cheng review the problem} in Section~\ref{sec:problem}, {\cheng review the state-of-the-art algorithm \texttt{Quick+} in Section~\ref{sec:quick-plus}, review the related work in Section~\ref{sec:related} and conclude the paper in Section~\ref{sec:conclusion}.}

\section{Problems}
\label{sec:problem}

\begin{figure}[t]
	\centering
	\includegraphics[width=0.80\linewidth]{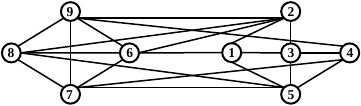}
        \vspace{-0.10in}
        \caption{Input graph used throughout the paper {\cheng (number $i$ represents vertex $v_i$)}}
	\label{fig:example graph}
\end{figure}

\subsection{Problem Definition}
\label{subsec:problem-definition}
We consider an undirected and unweighted graph $G=(V,E)$, where $V$ and $E$ are sets of vertices and edges respectively. 
Let $n$ be the number of vertices, i.e., $n = |V|$.
Given $H\subseteq V$, we use $G[H]$ to denote the subgraph of $G$ induced by $H$, i.e., $G[H]$ includes the set of vertices $H$ and the set of edges $\{(u,v)\in E \mid u,v \in H\}$. All subgraphs considered in this paper are \emph{induced} subgraphs.

Given $v\in V$, {\cheng we} let $\Gamma(v,V)$ (resp. $\overline{\Gamma}(v,V)$) denote the set of neighbours (resp. non-neighbours) of $v$ in $V$, i.e., $\Gamma(v,V)=\{u\in V\mid (u,v)\in E\}$ (resp. $\overline{\Gamma}(v,V)=\{u\in V\mid (u,v)\notin E\}$). We further define $\delta(v,V)=|\Gamma(v,V)|$ and $\overline{\delta}(v,V)=|\overline{\Gamma}(v,V)|$. We denote by $d$ the maximum degree of {\cheng a} vertex in $G$.

We then revisit the definition of $\gamma$-quasi-clique~\cite{pei2005mining,jiang2009mining,zeng2006coherent,liu2008effective,khalil2022parallel}.
\begin{definition}[$\gamma$-quasi-clique~\cite{pei2005mining}]
    Given $H\subseteq V$ and a fraction $0\leq \gamma \leq 1$, an induced subgraph $G[H]$ is said to be a $\gamma$-quasi-clique if and only if (1) $G[H]$ is connected and (2) for any vertex $v\in H$, {\CHENG it connects at least a fraction $\gamma$ of the vertices in $H$ (excluding $v$), i.e.,} $\delta(v,H)\geq \lceil\gamma\cdot (|H|-1)\rceil$.
\end{definition}

In particular, 
a 1-quasi-clique would reduce to a clique.
Besides, $\gamma$-quasi-clique has the following {\cheng two} properties. 

\begin{property}[Non-hereditary]
    \label{property-non-hereditary}
    For a $\gamma$-quasi-clique $G[H]$, a subgraph of $G[H]$ might not be a $\gamma$-quasi-clique. 
\end{property}

 This can be easily verified by an example in Figure~\ref{fig:example graph} where $G[\{v_1,v_3,v_4,v_5\}]$ is a 0.6-QC since each vertex connects at least 2 {\cheng out of 3 other} vertices whereas a subgraph $G[\{v_1,v_3,v_4\}]$ is not.

\begin{property}[2-diameter~\cite{pei2005mining}]
\label{property:diameter}
    For $\gamma\geq 0.5$, the diameter of a $\gamma$-quasi-clique is at most 2. 
\end{property}

{\cheng Following~\cite{pei2005mining}, we focus on those $\gamma$-quasi-cliques with $\gamma\geq 0.5$ only in this paper.}
This is because for a smaller value of $\gamma$, there exist numerous $\gamma$-quasi-cliques yet the majority of them are of small size and not cohesive~\cite{khalil2022parallel,guo2020scalable,pei2005mining}.
Moreover, 
prior studies~\cite{zeng2006coherent,liu2008effective,guo2020scalable,khalil2022parallel} often {\cheng focus on} a compact representation of the set of $\gamma$-quasi-cliques, namely the set of \emph{maximal} $\gamma$-quasi-cliques.

\begin{definition}[Maximal $\gamma$-quasi-clique]
    A $\gamma$-quasi-clique $G[H]$ is said to be maximal if and only if there is no other $\gamma$-quasi-clique $G[H']$ containing $G[H]$, i.e., $H\subseteq H'$.
\end{definition}

In this paper, we use QC (resp. MQC) as a shorthand of $\gamma$-quasi-clique (resp. maximal $\gamma$-quasi-clique) when the context is clear. Following~\cite{pei2005mining,liu2008effective,guo2020scalable,khalil2022parallel}, we consider a size threshold $\theta$ {\cheng for} each MQC $G[H]$ to be enumerated, namely $|H|\geq \theta$, since small MQCs are numerous and not statistically significant for real applications~\cite{khalil2022parallel,guo2020scalable,pei2005mining}.
Finally, we formalize the problem {\chengrr of enumerating MQCs with the size at least a threshold, which we call \emph{large} MQCs}.

\begin{problem}[Maximal $\gamma$-quasi-clique Enumeration~\cite{pei2005mining,liu2008effective,guo2020scalable,khalil2022parallel}]
\label{problem:MQCE}
    Given a graph $G=(V,E)$, a fraction threshold $\gamma\in [0.5, 1]$ and a positive integer size threshold $\theta$, the Maximal $\gamma$-Quasi-Clique Enumeration ({\cheng MQCE}) Problem aims to find all MQCs $G[H]$ {\cheng with} $|H|\geq \theta$.
\end{problem}

\noindent\textbf{NP-hardness.} The MQCE problem is NP-hard since the optimization problem {\cheng of} finding the MQC with the largest number of vertices is NP-hard~\cite{pastukhov2018maximum}. %
{\cheng Note that the optimization problem} can be solved by enumerating all MQCs and returning the largest one. 
{\CHENG Furthermore, } determining whether a QC is maximal is NP-hard~\cite{sanei2018enumerating}. 
{\CHENG In contrast, } determining if {\CHENG a structure that satisfies the hereditary property}, e.g., a clique~\cite{bron1973algorithm}, is maximal can usually be done in polynomial.

\subsection{Problem Decomposition}
\label{subsec:problem-decomposition}

{\revision
One common practice of enumerating {\chengrr large} maximal QCs is to (1) find a set of QCs that contains all maximal QCs, which may involve non-maximal QCs, and then (2) filter out non-maximal QCs with a post-processing procedure~\cite{liu2008effective,guo2020scalable,guo2022maximal}. 
This is mainly because checking whether a QC is maximal directly is NP-hard~\cite{sanei2018enumerating}.
{\chengr Therefore, we decompose the MQCE problem into two sub-problems, namely MQCE-S1 and MQCE-S2, each for a step involved in solving the MQCE problem, as follows.}

\begin{subproblem}[MQCE-S1]
\label{subproblem:SMQCE}
Given a graph $G=(V,E)$, a fraction threshold $\gamma\in [0.5, 1]$ and a positive integer size threshold $\theta$, 
{\chengr the MQCE-S1 problem is}
to find a set of QCs that contains all MQCs $G[H]$ with $|H|\geq \theta$.
\end{subproblem}

\begin{subproblem}[MQCE-S2]
\label{subproblem:NMQCF}
    Given a set $\mathcal{S}$ of QCs, 
    {\chengr the MQCE-S2 problem is to}
    filter out those that are subsets of others in {\chengr $\mathcal{S}$} and then return the remaining QCs.
    %
\end{subproblem}

The MQCE-S1 problem is \emph{NP-hard} since the problem of finding the largest MQC (which is NP-hard~\cite{pastukhov2018maximum}) can be solved by finding a set of QCs containing all MQCs and returning the largest one. 
{\chengr For the MQCE-S2 problem, we note that it is different from} the problem of determining whether a QC is maximal (which is NP-hard~\cite{sanei2018enumerating}).
For the former, the input is a set of QCs only. 
For the latter, the inputs include a QC and an input graph and the problem is to check whether there exists a superset of the QC in the input graph, which is also a QC. 
In fact, the {\chengr MQCE-S2} problem 
{\chengr can be solved in \emph{polynomial time}}
with respect to the size of input, {\chengr which we explain as follow}. 

{\chengr The MQCE-S2 problem is closely related to the \emph{set containment query}}, which is a fundamental problem in both database systems and theory of computer science~\cite{rivest1976partial,charikar2002new,bevc2009using,savnik2013index,savnik2021data}.   
%
Given a set of sets $\mathcal{S}$ 
and a query set $H$ of symbols from some alphabet,
{\chengr one type of set containment query called \texttt{GetAllSubsets} is to find all subsets of $H$ from $\mathcal{S}$.} 
{\chengr The state-of-the-art algorithm for \texttt{GetAllSubsets} can answer the query in $O(\min\{|\mathcal{S}|\cdot |H|, 2^{|H|}\})$ time with a set-trie data structure that can be built in $O(|\mathcal{S}|\cdot |H_{max}|)$ time, where $H_{max}$ is the largest set in $\mathcal{S}$~\cite{savnik2021data}.}
%
%

{\chengr Specifically, the MQCE-S2 problem} can be solved by iteratively 
{\chengr issuing a \texttt{GetAllSubsets} query for a QC $H$ in $\mathcal{S}$ and removing the found QCs from $\mathcal{S}$, which has been adopted by existing studies of enumerating MQCs~\cite{khalil2022parallel,guo2020scalable,liu2008effective}.}
%
{\chengr Consequently, the time complexity of this method is 
$O(\min\{|\mathcal{S}|^2\cdot \omega,~~|\mathcal{S}|\cdot 2^{2\omega}\})$ (which is polynomial wrt $|\mathcal{S}|$), where $\omega$ is the degeneracy of the input graph $G$. Note that the size of {\chengrr a $\gamma$-QC $H$, i.e., $|H|$,} is at most $2\omega+1$ for $\gamma\geq 0.5$
\ifx \CR\undefined
 
\else
~\cite{TR} 
\fi
and the cost of constructing the set-trie structure is dominated by that of issuing the \texttt{GetAllSubsets} query $O(|\mathcal{S}|)$ times.}



{\chengr We note that the time cost for solving the MQCE-S2 problem with the aforementioned method is typically small in practice due to the following reasons:} (1) we are usually interested in large MQCs only and 
{\chengr there are usually not many large QCs,}
i.e., $|\mathcal{S}|$ {\chengr is usually small} (see {\chengr the experimental results in} Table~\ref{tab:dataset}); and (2) we usually have $\omega<<n$ for the most real datasets which are sparse {\chengr (see the experimental results in Table~\ref{tab:dataset})}.  
For example, 
{\chengr the time cost of solving MQCE-S2 is}
within 0.1s for the majority of datasets and within 10s on all datasets we have used 
\ifx \CR\undefined
(as shown in the appendix).
\else
(as shown in the technical report~\cite{TR}).
\fi
{\chengr Therefore, in this paper, we focus on the MQCE-S1 problem, i.e., the one of finding a set of QCs containing all maximal QCs, as existing studies~\cite{liu2008effective,guo2020scalable,khalil2022parallel} did.}
}

 \section{The state-of-the-art branch-and-bound algorithm: \texttt{Quick+}}
\label{sec:quick-plus}

{\cheng In this part, we establish necessary background of branch-and-bound (BB) algorithms for {\revision {\chengr the MQCE-S1 problem, i.e., the one of} finding the set containing all MQCs} by reviewing the state-of-the-art BB algorithm, namely \texttt{Quick+}.}
{\cheng Specifically, \texttt{Quick+}} recursively partitions the search space (i.e., the set of possible vertex sets) to multiple sub-spaces via \emph{branching}. Each sub-space, {\cheng which corresponds to} a \emph{branch}, is represented by a triple of three vertex sets $(S,C,D)$ {\cheng explained} as follows. 
\begin{itemize}[leftmargin=*]
    \item \textbf{Partial set} $S$. Set of vertices that \emph{must} be included in every vertex set within the branch. 
    \item \textbf{Candidate set} $C$. Set of vertices that \emph{may} be included in $S$ in order to form larger vertex sets within the branch.
    \item \textbf{Exclusion set} $D$. Set of vertices that \emph{must not} be included in any vertex set within the branch.
\end{itemize}
{\cheng That is, each branch $(S,C,D)$ covers}
all those vertex sets that (1) include $S$ and (2) are subgraphs of $G[S\cup C]$.

Specifically, \texttt{Quick+} starts from the universal search space {\CHENG $(S, C, D)$} with $S=\emptyset$, $C=V$, and $D=\emptyset$, and recursively {\CHENG creates} branches 
{\cheng as follows.}
Consider a current branch $B=(S,C,D)$ with $C=\{v_1,v_2,...,v_{|C|}\}$. 
{\CHENG It creates $|C|$ branches, denoted by $B_i=(S_i,C_i,D_i)$ for $1\le i\le |C|$, from branch $B$.
{\cheng Branch $B_i$ covers} all vertex sets {\cheng that} include $S\cup \{v_i\}$ and exclude $D \cup \{v_1,...,v_{i-1}\}$.
}
%
Formally, for $1\leq i \leq |C|$, we have
\begin{equation}
    \label{eq:SE_branching}
    S_i = S \cup \{ v_i\};~~D_i = D \cup \{v_1, v_2, ..., v_{i-1}\};~~C_i = C - \{ v_1, v_2, ..., v_i\}
\end{equation}
{\cheng We call this branching the \emph{SE branching},  as illustrated in Figure~\ref{fig:illustration_branching}(a).

\begin{figure}[t]
	\centering
	\includegraphics[width=0.85\linewidth]{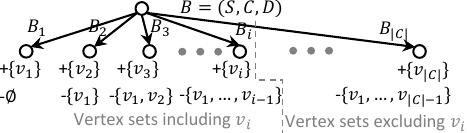}
	\\
	(a) SE Branching
	\vspace{0.10in}
	\\
	\includegraphics[width=0.85\linewidth]{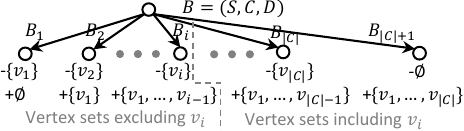}
	\\
	(b) Sym-SE Branching
	\vspace{-0.10in}
 	\caption{Illustration of SE {\cheng branching} and Sym-SE branching 
  {\cheng (``+'' means to \emph{include} a set of vertices, i.e., the set $S$ is expanded with these vertices; ``-'' means to \emph{exclude} a set of vertices, i.e., the set $D$ is expanded with these vertices)}
  }
	\label{fig:illustration_branching}
\end{figure}

We note that SE branching and the existing branching strategy adopted by the Bron-Kerbosch (BK) algorithm~\cite{bron1973algorithm}, which we call \emph{BK branching}, share the way of forming the branches. The difference is that BK branching is used for enumerating maximal subgraph structures that satisfy the hereditary property (e.g., cliques). Specifically, it would further prune some of the formed branches based on the hereditary property. In contrast, SE branching does not require the subgraphs to be enumerated to satisfy the hereditary property - this is why it is adopted by Quick+ for enumerating MQCs, and it cannot prune some formed branches as BK branching does.}

During the recursive branching process, \texttt{Quick+} applies two types of pruning techniques, namely Type I pruning {\cheng rules} and Type II pruning {\cheng rules}. 
Intuitively, Type I pruning rules are conducted on $C$ and aim to refine $C$ by removing those vertices that satisfy certain conditions; Type II pruning rules are conducted on $S$ and aim to prune those branches where vertices in $S$ satisfy certain conditions. 
The rationale behind is that if a vertex $v$ satisfies certain conditions, {\cheng each} MQC {\cheng covered by} this branch {\cheng does not} include this vertex, and thus we can either remove $v$ from {\cheng $C$ for} this branch, {\cheng i.e., Type I pruning rules apply} (if $v\in C$), or prune the entire branch, {\cheng i.e., Type II pruning rules apply} (if $v \in S$). For simplicity, we omit the details of these pruning techniques 
and refer them to ~\cite{khalil2022parallel}.

\begin{algorithm}{}
\small
\caption{{\cheng An existing branch-and-bound algorithm}: \texttt{Quick+}~\cite{khalil2022parallel}}
\label{alg:enumeration_scheme}
\KwIn{A graph $G=(V,E)$, $0.5 \leq \gamma\leq 1$, and $\theta>0$}
\KwOut{{\chengrr A set of QCs that includes} all MQCs}
\texttt{Quick-Rec}$(\emptyset,V,\emptyset)$;\\
\SetKwBlock{Enum}{Procedure \texttt{Quick-Rec}$(S,C, {\cheng D})$}{}
\Enum{
    \tcc{Termination}
    \If{$C=\emptyset$}{
        \If{$G[S]$ is a QC}{
            \textbf{Output} $G[S]$ if $|S|\geq \theta$; \textbf{return} true;
        }
        \textbf{return} false\;
    }
    \tcc{{\cheng SE} Branching}
    Create $|C|$ branches $B_i = (S_i, C_i, D_i)$ based on Equation~(\ref{eq:SE_branching});\\
    \For{each branch $B_i$}{
        \tcc{Pruning before the next recursion}
        $C_i'\leftarrow$ Type I pruning rules on $C_i$\;
        \lIf{any of Type II pruning on $S_i$ is triggered}{\textbf{continue}}
        $\mathcal{T}_{i}\leftarrow$ \texttt{Quick-Rec}$(S_i,C_i',D_i)$;
    }
    \tcc{Additional step: output $G[S]$ if necessary}
    \If{all of $\mathcal{T}_{i}$ are false}{
        \If{$G[S]$ is a QC}{
            \textbf{Output} $G[S]$ if $|S|\geq \theta$; \textbf{return} true;   
        }
        \textbf{return} false;
    }
    \textbf{return} true;
}
\end{algorithm}

We finally summarize \texttt{Quick+} in Algorithm~\ref{alg:enumeration_scheme}. Specifically, it starts from the branch $(B,C,D)=(\emptyset,V,\emptyset)$ {\cheng by calling a recursive procedure called \texttt{Quick-Rec} (line 1)}, recursively {\cheng creates} branches {\cheng via SE branching (line 7)}, and conducts the aforementioned pruning {\cheng operations (line 9-10)}. In particular, it terminates the branch once $C=\emptyset$, and outputs the partial set $G[S]$ only if $G[S]$ is a QC and $|S|\geq \theta$ (line 4-5). We remark that \texttt{Quick+} does not check whether an output QC is maximal or not (mainly due to its NP-hardness). Therefore, it would return a superset of all MQCs which inevitably contains some non-maximal QCs, 
{\revision {\chengr i.e., it solves the MQCE-S1 problem, but not the MQCE problem.}} 

Besides, {\CHENG we note that for a branch $(S, C, D)$,} $G[S]$ could be a MQC {\CHENG even} if {\CHENG no} QCs are found in the created sub-branches {\CHENG due to the non-hereditary property of QC}.
{\CHENG Therefore, \texttt{Quick-Rec} monitors whether a sub-branch of the current one would find a QC. If so, it returns true (e.g., line 5, line 14 and line 16); if not, it returns false (e.g., line 6 and line 15). In the case that a QC is found in a sub-branch, the QC should be a superset of $G[S]$, i.e., $G[S]$ cannot be a MQC, and therefore, there is no need to consider $G[S]$. In the other case that no QCs are found in any of the sub-branches (line 12), it checks if $G[S]$ is a large QC and outputs it if so (line 13-14).}

\smallskip
\noindent\textbf{Time Complexity}. \texttt{Quick+} would explore $O(2^{n})$ branches in the worst case, though some pruning rules are applied to boost its performance {\cheng in practice}. Hence, the worst-case time complexity is $O^*(2^{n})$, where $O^*$ suppresses the polynomials~\cite{khalil2022parallel}. 

\section{A New Branch-and-Bound Algorithm: \texttt{FastQC}}
\label{sec:FastQC}
{\cheng 
In this section, we introduce our new branch-and-bound (BB) algorithm called \texttt{FastQC}.
First, we develop a new pruning technique, which is based on a necessary condition for a branch to hold QCs, {\cheng i.e., a vertex set within the branch corresponds to a QC} (Section~\ref{subsec:necessary_condition}), and introduce a method to apply the pruning technique in a \emph{progressive} fashion by refining a branch and re-checking the necessary condition iteratively (Section~\ref{subsec:progressively_RR}).
Second, we observe that the pruning technique has a nice property that if a branch with a partial set $S$ can be pruned, then any branch with the partial set as a \emph{superset} of $S$ can be pruned as well. To better utilize this property, we adopt a new branching method that is \emph{symmetric} to SE branching, which we call \emph{Sym-SE branching} (Section~\ref{subsec:Sym-PivSE}). The rationale is that Sym-SE branching would produce a series of branches such that the following branches have their partial sets as \emph{supersets} of those of preceding ones. As a result, if we find a branch that can be pruned, we can prune all branches following this branch in the series safely. Third, we observe that in certain cases, SE branching and Sym-SE branching can be \emph{jointly} applied so that more branches can be pruned. We call the resulting branching method the \emph{Hybrid-SE branching} and present it in Section~\ref{subsec:hybrid-SE}.
Finally, we summarize the \texttt{FastQC} algorithm, which is a BB algorithm based on the newly developed pruning techniques and branching methods and analyze its time complexity in Section~\ref{subsec:summary_fastqc}. In particular, \texttt{FastQC} has the worst-case time $O(n\cdot d\cdot \alpha_k^{n})$ {\chengrr with} $\alpha_k< 2$.
}

\subsection{A Novel Necessary Condition for a Branch To Hold QCs}
\label{subsec:necessary_condition}

%
{\cheng Consider a branch $B=(S,C,D)$. We aim to find an easily tractable \emph{necessary condition for a current branch to hold QCs}~\footnote{We note that the problem of determining whether branch $B$ holds a QC (or formally, whether one of the partial sets of the branches under $B$ induces a QC) is hard. In fact, we prove that this problem is NP-hard and for simplicity, we put the proof in the 
\ifx \CR\undefined
appendix
\else
technical report~\cite{TR}
\fi
.}.}
Then, for those branches that violate the condition, they hold no QCs and thus can be pruned safely.
%
{\cheng We will show that with this pruning technique employed (and some branching method designed accordingly), we would need to explore strictly fewer than $O(2^{n})$ branches in theory.} 
Below, we give the details of the condition.

\begin{figure}[]
	\centering
	\begin{tabular}{c c}
		\begin{minipage}{3.80cm}
			\includegraphics[width=3.8cm]{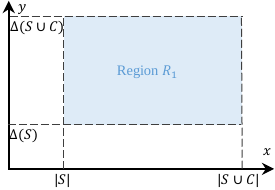}
		\end{minipage}
		&
		\begin{minipage}{3.80cm}
			\includegraphics[width=3.8cm]{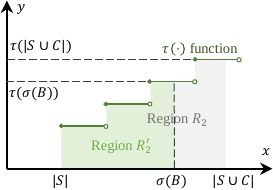}
		\end{minipage}		
		\\
		(a) Region $R_1$
		&
		(b) Region $R_2$ and Region $R_2'$ 
	\end{tabular}
	\vspace{-0.10in}
	\caption{{\cheng Illustration of} Condition C1 (Region $R_1$) and Condition C2 (Region $R_2$ and Region $R_2'$) in the SD space}
	\label{fig:SD_Exp1}
\end{figure}

{\cheng Let $H$ be a set of vertices.} We define $\Delta(H)$ {\cheng to be} the maximum number of disconnections of {\cheng a} vertex in $H$ {\cheng within $G[H]$}. Formally, we have
\begin{equation}
    \Delta(H)=\max_{v\in H} \overline{\delta}(v, H).
\end{equation}
Consider a subset {\cheng $H_{sub}$} and a superset {\cheng $H_{sup}$} of $H$, we have 
\begin{equation}
\label{eq:delta}
    \Delta(H_{sub}) \leq \Delta(H) \leq \Delta(H_{sup})\  \text{for}\ H_{sub}\subseteq H \subseteq H_{sup}.
\end{equation}
This can be verified by {\cheng the fact that the set of} disconnections within a subgraph $G[H_{sub}]$ (resp. a supergraph $G[H_{sup}]$) is always a subset (resp. a superset) of that within $G[H]$. 

{\cheng Given a graph $G[H]$, we map it to a 2-dimensional space at the point $(|H|, \Delta(H))$. We call this space the \emph{\underline{s}ize \underline{d}isconnection space} (SD space). 
We note that a point $(x,y)$ in the SD space corresponds to a set of possible graphs $G[H]$ with $|H|=x$ and $\Delta(H)=y$. Note that we can focus on the first quadrant of the SD space, namely $x\geq 0$ and $y\geq 0$. 

With the defined SD space, we proceed to introduce two necessary conditions for a branch $B$ to hold QCs, namely C1 and C2.

\smallskip
\noindent\textbf{Condition C1.} 
For a QC $G[H]$ under the branch $B$, its point in the SD space must reside in a rectangular region $R_1$ defined as follows.
\begin{equation}
\label{eq:condition1}
    \text{Region $R_1$: } |S| \leq x\leq |S\cup C| \text{  and  } \Delta(S) \leq y \leq \Delta(S\cup C) 
\end{equation}
This is because $S\subseteq H \subseteq S\cup C$ and thus $\Delta(S)\leq \Delta(H)\leq \Delta(S\cup C)$ according to Equation~(\ref{eq:delta}). 
An illustration of Region $R_1$ is shown in Figure~\ref{fig:SD_Exp1} (a) {\cheng (the blue region)}.
Correspondingly, we obtain the following necessary condition.

\medskip
\noindent\fbox{%
    \parbox{0.47\textwidth}{%
       \textbf{Condition C1:} If a branch $B$ holds a QC $G[H]$, then the point of $G[H]$ {\cheng in the SD space} resides in Region $R_1$.
    }%
}

\smallskip
\noindent\textbf{Condition C2.} 
Recall that for a QC $G[H]$, each vertex $v$ in $G[H]$ has the number of its connections within $G[H]$ \emph{at least} $\lceil\gamma \cdot (|H| - 1)\rceil$ by definition, i.e., 
\begin{equation}
\forall v,~~\delta(v, H) \ge \lceil\gamma \cdot (|H| - 1)\rceil
\label{eq:connections} 
\end{equation}
Equivalently, each vertex $v$ has the number of its disconnections within $G[H]$, which is equal to $|H| - \delta(v, H)$, \emph{at most} $|H| - \lceil\gamma \cdot (|H| - 1)\rceil$, i.e., 
\begin{equation}
\forall v,~~\overline{\delta}(v, H) \le |H| - \lceil\gamma \cdot (|H| - 1)\rceil = \lfloor(1-\gamma) \cdot |H| + \gamma\rfloor
\label{eq:disconnections} 
\end{equation}
Equation~(\ref{eq:disconnections}) implies that the maximum number of disconnections of a vertex in $G[H]$ is also \emph{at most} $\lfloor(1-\gamma) \cdot |H| + \gamma\rfloor$, i.e., 
\begin{equation}
\Delta(H) \le \lfloor(1-\gamma) \cdot |H| + \gamma\rfloor
\label{eq:max-disconnections} 
\end{equation}
Equation~(\ref{eq:max-disconnections}) essentially says that the point of any QC $G[H]$ in the SD space is \emph{below} the curve representing the following function $\tau(x)$ {\cheng inclusively}.
\begin{equation}
\tau(x) := \lfloor(1-\gamma) \cdot x + \gamma\rfloor
\label{eq:tau-x} 
\end{equation}
We note that $\tau(x)$ corresponds to a piece-wise {\cheng and non-decreasing} function and each piece is a horizontal line segment with the left endpoint being included and the right endpoint being excluded. An illustration of $\tau(x)$ is in Figure~\ref{fig:SD_Exp1}(b).
Based on Equation~(\ref{eq:max-disconnections}) and Equation~(\ref{eq:tau-x}), we deduce the following lemma.

\begin{lemma}
\label{lemma:TD_Property2}
    A graph $G[H]$ {\cheng is} a QC iff  $\Delta(H) \leq \tau (|H|) $.
\end{lemma}

Based on Lemma~\ref{lemma:TD_Property2} and the fact that all partial sets within $B$ have the size between $|S|$ and $|S\cup C|$, we deduce that 
the points of those QCs within $B$ (if any)
must reside in a region $R_2$ as defined below.
\begin{equation}
\label{eq:condition2-}
   \text{Region $R_2$: }  |S| \leq x \leq |S\cup C| \text{ and }  0 \leq y \leq \tau(x).
\end{equation}
An illustration of Region $R_2$ is shown in Figure~\ref{fig:SD_Exp1}(b) {\cheng (the grey and green region)}.

We note that the upper bound $|S\cup C|$ of the size of a QC under branch $B$ can often be loose. We therefore tighten it to be $\sigma(B)$, which is defined as follows.
\begin{equation}
\label{eq:fun_sigma}
\sigma(B) =
\left\{
\begin{array}{ll}
      |S \cup C| & S = \emptyset \\
      \min\{|S \cup C|, d_{min}(B)/\gamma + 1\} & S \neq \emptyset \\
\end{array} 
\right. 
\end{equation}
where $d_{min}(B)$ is the minimum degree of a vertex in $S$ within $G[S\cup C]$.
That is, 
\begin{equation}
    d_{min}(B) = \min_{v\in S}\delta(v, S \cup C)
    \label{eq:d-min}
\end{equation}
We verify that for a QC $G[H]$ under branch $B$ (if any), we have $|H| \le \sigma(B)$, which we formally present in the following lemma (the proof is put in the 
\ifx \CR\undefined
appendix for simplicity.
\else
technical report~\cite{TR} for simplicity).
\fi
\begin{lemma}
For any QC $G[H]$ under branch $B$, we have $|H| \le \sigma(B)$.
\label{lemma:tight_bound}
\end{lemma}
Based on Lemma~\ref{lemma:tight_bound}, we obtain a region $R_2'$ as defined below, which covers all possible QCs under branch $B$ and is narrower than $R_2$.
\begin{equation}
\label{eq:condition2}
  \text{Region $R_2'$: }   |S| \leq x \leq \sigma(B) \text{ and } 0 \leq y \leq \tau(x).
\end{equation}
An illustration of Region $R_2'$ is shown in Figure~\ref{fig:SD_Exp1}(b) {\cheng (the green region)}.
Correspondingly, we obtain the following necessary condition.

\medskip
\noindent\fbox{%
    \parbox{0.47\textwidth}{%
       \textbf{Condition C2:} If a branch $B$ holds a QC $G[H]$, then the point of $G[H]$ {\cheng in the SD space} resides in Region $R_2'$.
    }%
}
\medskip

We note that in the case that $\sigma(B) < |S|$, it means that Region $R_2'$ is empty, which implies that Condition C2 is not satisfied and there exist no QCs under the branch $B$.
}

\begin{figure}[]
	\centering
	\begin{tabular}{c c}
		
		\begin{minipage}{3.80cm}
			\includegraphics[width=3.8cm]{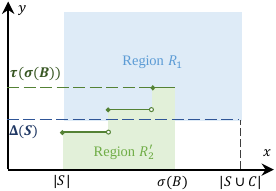}
		\end{minipage}	
        &
        \begin{minipage}{3.80cm}
			\includegraphics[width=3.8cm]{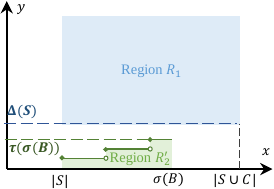}
		\end{minipage}
		\\
        {\cheng (a) C1\&2 is satisfied}
		&
        {\cheng (b) C1\&2 is not satisfied}
	\end{tabular}
	\vspace{-0.10in}
	\caption{{\cheng Illustration of the} necessary condition in the SD space}
	\label{fig:SD_Exp2}
\end{figure}

{\cheng
\smallskip
\noindent\textbf{Necessary Condition (Summary).} In summary, if a branch $B$ holds a QC $G[H]$, then the point of $G[H]$ in the SD space must reside in both Region $R_1$ and Region $R_2'$.
It further implies that the intersection of the two regions, which we denote by $R_{1\&2} = R_1 \cap R_2'$, is non-empty. We use this as the necessary condition for branch $B$ to hold a QC and shall show that it can be verified efficiently in $O(d)$ time. 
Specifically, we have the following necessary condition.

\medskip
\noindent\fbox{%
    \parbox{0.47\textwidth}{%
       \textbf{Condition C1\&2:} If a branch $B$ holds a QC, then $R_{1\&2} = R_1 \cap R_2'$ is non-empty.
    }%
}
\medskip

For illustration, we show the case where the necessary condition C1\&2 is satisfied in Figure~\ref{fig:SD_Exp2}(a) and the case where C1\&2 is not satisfied in Figure~\ref{fig:SD_Exp2}(b). 

We notice that the necessary condition C1\&2, i.e., Region $R_{1\&2}$ is non-empty, is equivalent to that $\Delta(S) \leq \tau(\sigma(B))$. This is because when $\Delta(S) \leq \tau(\sigma(B))$, regions $R_1$ and $R_2'$ would intersect and when $\Delta(S) > \tau(\sigma(B))$, the regions would not intersect, and vice versa, as illustrated in Figure~\ref{fig:SD_Exp2}. 
Correspondingly, we have the following equivalent necessary condition.

\medskip
\noindent\fbox{%
    \parbox{0.47\textwidth}{%
       \textbf{Condition C1\&2 (equivalent):} If a branch $B$ holds a QC, then $\Delta(S) \leq \tau(\sigma(B))$.
    }%
}

\smallskip
\noindent\textbf{Time Complexity of Checking the Condition C1\&2.} 
The cost is dominated by the that of computing $\Delta(S)$ and $d_{min}(B)$. First, we can maintain two arrays to record the degree of each vertex $v$ within $G[S]$ (i.e., $\delta(v,S)$) and that within $G[S\cup C]$ (i.e., $\delta(v,S\cup C)$). The maintenance cost is $O(d)$ since when forming a branch $B_i$ by including a vertex $v_i$ to $S$ (recall Equation~(\ref{eq:SE_branching})), we only need to update $\delta(\cdot,S)$ and $\delta(\cdot,S\cup C)$ for $v_i$'s neighbours and there are $O(d)$ neighbors). Second, we can compute $\Delta(S)$ by scanning the vertices in $S$ and their degrees within $G[S]$ (which have been maintained), and the cost is $O(|S|)$. Similarly, we can compute $d_{min}(B)$ by scanning the vertices in $S$ and their degrees within $G[S\cup C]$ (which have been maintained), and the cost is $O(|S|)$. Third, for a non-empty $S$, we have $|S|\leq \sigma(B)$ based on Lemma~\ref{lemma:tight_bound} and  $\sigma(B)\leq d/\gamma+1 \leq 2d+1$ based on Equation~(\ref{eq:fun_sigma}) and $\gamma\geq 0.5$. In summary, the cost of checking C1\&2 is $O(d) + O(|S|) + O(|S|) = O(d)$.

}

{\color{blue}
}

{\cheng 
\subsection{Progressively Refining a Branch and Re-Checking the Necessary Condition}
\label{subsec:progressively_RR}
Consider a branch $B = (S, C, D)$. We first check if the necessary condition C1\&2, i.e., $\Delta(S) \le \tau(\sigma(B))$, is satisfied. If no, we can prune $B$ directly; If yes, while we cannot prune $B$ immediately, we may be able to \emph{refine} $B$ with the information $\tau(\sigma(B))$ by removing some vertices from $C$. We then \emph{re-check} the condition for the refined branch, which we denote by $B'$, and prune it if the condition is not satisfied. The rationale is that with some vertices removed from the candidate set $C$, the necessary condition would become less likely to be satisfied and correspondingly the branch can be pruned more likely, as will be explained later. In fact, we can repeat this process until either (1) the branch is pruned; or (2) the branch cannot be refined further. We provide the details as follows.

\smallskip
\noindent\textbf{Refining a Branch.} With the information of $\tau(\sigma(B))$, we can possibly refine $B$ by removing from $C$ some vertices as follows.
\begin{itemize}[leftmargin=*]
    \item \textbf{Refinement Rule 1.} Remove from $C$ those vertices $v$ with $\Delta(S\cup\{v\})> \tau(\sigma(B))$
    \item \textbf{Refinement Rule 2.} Remove from $C$ those vertices $v$ with $\delta(v,S\cup C)< \theta-\tau(\sigma(B))$
\end{itemize}
For Rule (1), it is because any QC $G[H]$ under branch $B$ cannot hold vertex $v$ since otherwise we deduce that
$\Delta(S\cup\{v\})\leq \Delta(H)\leq \tau(|H|)\leq \tau(\sigma(B))$, which contradicts to $\Delta(S\cup \{v\})>\tau(\sigma(B))$.
Here, $\Delta(S\cup\{v\})\leq \Delta(H)$ is because $S\cup \{v\} \subseteq H$ (by assumption), $\Delta(H)\leq \tau(|H|)$ is because $G[H]$ is a QC, and $\tau(|H|)\leq \tau(\sigma(B))$ is because $|H| \leq \sigma(B)$ (based on Lemma~\ref{lemma:tight_bound}) and $\tau(\cdot)$ is non-decreasing.
For Rule (2), {\cheng it is because} only those QCs with the size at least $\theta$ are to be found. For each vertex $v$ in such a large QC $G[H]$ under branch $B$, we have $\delta(v, S\cup C)\geq \delta(v, H) = |H|- \overline{\delta}(v,H)\geq |H|-\Delta(H)\geq |H|-\tau(|H|)\geq |H|- \tau(\sigma(B))\geq \theta-\tau(\sigma(B))$.

\smallskip
\noindent\textbf{Re-checking the Necessary Condition.} Suppose the branch $B = (S, C, D)$ has been refined to $B' = (S, C', D)$ with $C' \subset C$. We note that $\tau(\sigma(B'))$ would be possibly smaller than $\tau(\sigma(B))$. This is because $C' \subset C$ implies $d_{min}(B') \le d_{min}(B)$, which further implies $\sigma(B') \le \sigma(B)$, and $\tau(\cdot)$ is non-decreasing. Therefore, we can re-check the necessary condition for branch $B'$, i.e., $\Delta(S) \le \tau(\sigma(B'))$, which is less likely to be satisfied than that for branch $B$, i.e., $\Delta(S) \le \tau(\sigma(B))$, given that $\tau(\sigma(B')) \le \tau(\sigma(B))$. If the condition is not satisfied, we prune branch $B'$.

\smallskip
\noindent\textbf{Repeated Process and Stopping Criterion.} In the case that the refined branch $B'$ cannot be pruned. We can repeat the process of refining $B'$ (based on the information of $\tau(\sigma(B'))$) and re-checking the condition for the refined branch. We stop the process until either (1) a refined branch is pruned or (2) a branch cannot be refined any further (i.e., no vertices can be removed from the candidate set).
}

For illustration, consider the problem of finding all 0.7-MQCs in Figure~\ref{fig:example graph} and a branch $B$ with $S=\{v_1,v_3,v_4\}$, $C=\{v_2,v_5,v_6,v_7,v_8,v_9\}$ and $D=\emptyset$. 
%
{\cheng First, we check necessary condition for $B$. Specifically, we compute $\Delta(S) = \overline{\delta}(v_1,S) =|\{v_1,v_4\}| = 2$}, 
$\sigma(B)=\min\{9,4/0.7+1\}=6.71$ and $\tau(\sigma(B))=\tau(6.71)=\lfloor 0.3\times 6.71 +0.7 \rfloor=2$.
{\cheng Since $\Delta(S) = 2 \leq \tau(\sigma(B)) = 2$, i.e., the necessary condition for $B$ is satisfied, we cannot prune $B$.
Then, we refine $B$ by removing vertices $v_6,v_7,v_8,v_9$ from $C$ since for each such vertex $v$, we have $\Delta(S\cup \{v\})> \tau(\sigma(B)) = 2$ (i.e., Refinement Rule 1 applies).
%
We denote the refined branch as $B'$ and re-check the necessary condition for $B'$. Specifically, we compute $\sigma(B')=\min\{5,2/0.7+1\}=3.85$ and $\tau(\sigma(B'))=\tau(3.85)=1$. Since $\Delta(S) = 2 > \tau(\sigma(B')) = 1$, i.e., the necessary condition for $B'$ is not satisfied, we can safely prune $B'$ and stop the process.}

\begin{figure}[]
	\centering
	\begin{tabular}{c c}
		\begin{minipage}{3.80cm}
			\includegraphics[width=4.1cm]{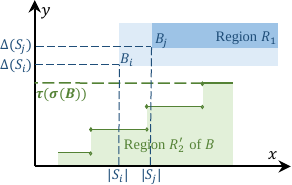}
		\end{minipage}
        &
		\begin{minipage}{3.80cm}
			\includegraphics[width=4.1cm]{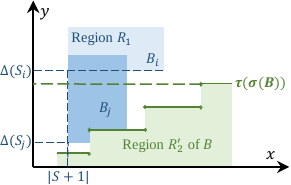}
		\end{minipage}	
		\\
        {(a) Sym-SE branching}
		&
        {(b) SE branching}
	\end{tabular}
	\vspace{-0.10in}
	\caption{{\cheng Illustration of} branching at $B$ in the SD space}
	\label{fig:SD_branching}
\end{figure}

\subsection{A {\cheng Symmetric} Branching Strategy {\cheng of SE Branching}: {\cheng Sym-SE Branching}}
\label{subsec:Sym-PivSE}

Consider a branch $B=(S,C,D)$, {\cheng which} satisfies the necessary condition C1\&2, i.e., $\Delta(S)\leq \tau(\sigma(B))$, after the 
{\cheng the progressive process of refining a branch and re-checking the necessary condition.}
We need to create sub-branches from branch $B$ via branching. 
{\cheng One option is to adopt the SE branching, which is defined in Equation~(\ref{eq:SE_branching}), as Quick+ does. Another option is to adopt a branching {\cheng that is \emph{symmetric} to} the SE branching, called \emph{Sym-SE branching}, which is defined as follows.
It creates $(|C|+1)$ sub-branches from $B$, namely $B_1, B_2, ..., B_{|C|+1}$. Branch $B_i$ ($1\le i\le |C|+1$) covers those vertex sets, each (1) including all vertices in $S \cup \{v_1,v_2,...,v_{i-1}\}$; (2) excluding all vertices in $D \cup \{v_i\}$. Formally, branch $B_i$ is defined as follows.
\begin{equation}
    S_i\!=\!S\cup \{v_1,v_2,...,v_{i-1}\},\ D_i\!=\!D\cup\{v_i\},\ C_i\!=\!C-\{v_1,v_2,...,v_i\}.
\label{eq:sym-se}
\end{equation}
Here, $v_0$ in branch $B_1$ and $v_{|C|+1}$ in branch $B_{|C|+1}$ are both fictitious.
An illustration of the Sym-SE branching is shown in Figure~\ref{fig:illustration_branching}(b).
{\cheng The two branching methods are symmetric because as shown in Figure~\ref{fig:illustration_branching}, (1) for SE branching, branches $B_1$, ..., $B_i$ \emph{include} vertex $v_i$ while the remaining branches \emph{exclude} vertex $v_i$ and (2) for Sym-SE branching, branches $B_1$, ..., $B_i$ \emph{exclude} vertex $v_i$ while the remaining branches \emph{include} vertex $v_i$.
}

We note that a symmetric branching of the BK branching, which is called \emph{Sym-BK branching}, has also been explored for enumerating maximal subgraphs that satisfy the hereditary property~\cite{yu2022maximum,zhou2020enumerating}. Sym-SE branching and Sym-BK branching share the way of forming branches. The difference is that Sym-BK branching can prune some of the formed branches based on the hereditary property. In contrast, Sym-SE does not require the subgraphs to be enumerated to satisfy the hereditary property, and correspondingly it cannot prune some formed branches as Sym-BK does.
%

We show that the necessary condition C1\&2 defined in the SD space would work more effectively when the Sym-SE branching is used than when the SE branching is used. The reason is two-fold. \underline{First}, a created branch $B_i$ by Sym-SE branching would have a larger chance to violate the necessary condition C1\&2, i.e., $\Delta(S_i) > \tau(\sigma(B))$~\footnote{{\cheng We note that for sub-branches, we use a looser condition than the necessary condition for branch $B_i$, i.e., $\Delta(S_i) {\cheng \le} \tau(\sigma(B_i))$. 
}}, and be pruned, when $i$ gets larger. This is due to the fact that $S_i$ involves $|S|+i-1$ vertices, and correspondingly we have $\Delta(S_i)$ increase with $i$. 
%
\underline{Second}, if a branch $B_i$ by Sym-SE branching violates the necessary condition (i.e., $\Delta(S_i)>\tau(\sigma(B))$) and can be pruned, then all other branches $B_j$ following $B_i$ with $j > i$ would violate the necessary condition (i.e., $\Delta(S_j)>\tau(\sigma(B))$) and can be pruned also. This is due to the fact that the partial set $S_j$ is always a superset of $S_i$ and thus we have $\Delta(S_j) \geq \Delta(S_i) > \tau(\sigma(B))$.

For illustration, consider Figure~\ref{fig:SD_branching}(a) for Sym-SE branching, where when a branch $B_i$ can be pruned, any branch $B_j$ with $j>i$ can also be pruned. Consider Figure~\ref{fig:SD_branching}(b) for SE branching, where a branch $B_i$ can be pruned does not imply that any branch $B_j$ with $j>i$ can be pruned.
}

%
%

{\cheng
\smallskip
\noindent\textbf{Ordering of Vertices in $C$.}
Sym-SE branching implicitly uses an ordering of vertices in $C$, which we specify as follows.
}
{\cheng We} find a smaller subset $C'$ of $C$ and put them before other vertices in the ordering such that including them to the partial set collectively would {\cheng cause} one sub-branch {\cheng to} violate the {\cheng necessary} condition (i.e., $\Delta(S\cup C')>\tau(\sigma(B))$) and thus {\cheng this branch and also the} following sub-branches can be pruned. Specifically, the ordering is defined based on a vertex $\hat{v}$ called \emph{pivot}, which is selected {\cheng from those vertices in $S\cup C$ that} have more than $\tau(\sigma(B))$ disconnections among $S\cup C$, i.e., $\overline{\delta}(\hat{v},S\cup C)>\tau(\sigma(B))$.
We define
\begin{equation}
    a=\tau(\sigma(B))-\overline{\delta}(\hat{v},S)\ \text{and}\ b=\overline{\delta}(\hat{v},C),
\end{equation}
where $a$ denotes the largest possible number of vertices that can be included from $\overline{\Gamma}(\hat{v},C)$ to $S$ without violating the necessary condition. Note that $a<b$ since $b-a=\overline{\delta}(\hat{v},S\cup C)-\tau(\sigma(B))>0$.
We put those vertices in $\overline{\Gamma}(\hat{v},C)$ before others in the ordering. There are two cases.

\noindent\underline{Case 1}: $\hat{v}\in S$. We define the ordering of vertices in $C$ as follows.
\begin{equation}
    \label{sym-pivse-case1}
    \langle v_1,v_2,...,v_b,v_{b+1},...,v_{|C|} \rangle,
\end{equation}
where the first $b$ vertices, namely $v_1,v_2,...,v_b$, are from $\overline{\Gamma}(\hat{v},C)$ in any order, and the others are from $\Gamma(\hat{v},C)$ in any order. Then, branch $B_{a+2}$ would violate the necessary condition since $\Delta(S_{a+2})\geq \overline{\delta}(\hat{v},S_{a+2})  = \overline{\delta}(\hat{v}, S\cup \{v_1, v_2, ..., v_{a+1}\}) = \overline{\delta}(\hat{v}, S) + \overline{\delta}(\hat{v}, \{v_1, v_2, ..., v_{a+1}\}) = \overline{\delta}(\hat{v},S)+a+1=\tau(\sigma(B))+1$. Consequently, the branches $B_{a+2}, B_{a+3},...,B_{|C|+1}$ violate the necessary condition and can be pruned. 

\begin{figure}[t]
	\centering
	\includegraphics[width=0.78\linewidth]{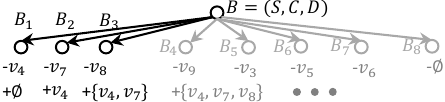}
	\\
	(a) Sym-SE branching {\cheng with $v_1$ as the pivot (Case 1)}
	\vspace{0.10in}
	\\
	\includegraphics[width=0.85\linewidth]{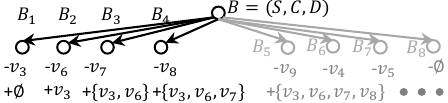}
	\\
	(b) Sym-SE branching: {\cheng with $v_3$ as the pivot (Case 2)}
	\vspace{0.10in}
	\\
	\includegraphics[width=0.85\linewidth]{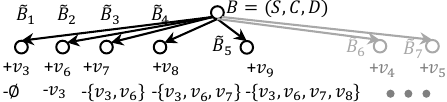}
	\\
	(c) SE branching {\cheng with $v_3$ as the pivot}
        \vspace{0.10in}
	\\
	\includegraphics[width=0.85\linewidth]{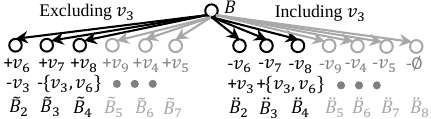}
	\\
	(d) Hybrid-SE branching {\cheng with $v_3$ as the pivot}
 	\vspace{-0.10in}
	\caption{Illustration of {\cheng Sym-SE branching ((a) and (b)), SE branching (c) and Hybrid-SE branching (d)} at $B=(S,C,D)$ with $S=\{v_1,v_2\}$, $C=\{v_3,v_4,...,v_9\}$ and $D=\emptyset$  ($\gamma=0.6$ and $\tau(\sigma(B))=3$)}
	\label{fig:branching_case}
\end{figure}

 {\cheng For illustration}, consider the example of finding all 0.6-MQCs from the graph given by Figure~\ref{fig:example graph} as shown in Figure~\ref{fig:branching_case}(a). Based on pivot $v_1$ in $S$, {\cheng which} disconnects 4 vertices in $C$, i.e., $\{v_4,v_7,v_8,v_9\}$, we define the ordering $\langle v_4,v_7,v_8,v_9,v_3,v_5,v_6\rangle$. The branches $B_4, ..., B_9$ can be pruned since $B_4$ has $S_4=\{v_1,v_2,v_4,v_7,v_8\}$ and $\Delta(S_{4})  = 4 >\tau(\sigma(B))  = 3$.

\noindent\underline{Case 2}: $\hat{v}\in C$. We define the ordering of vertices in $C$ as follows.
\begin{equation}
    \label{sym-pivse-case2}
    \langle \hat{v}, v_2, v_3,...,v_b, v_{b+1},..., v_{|C|} \rangle,
\end{equation}
where the first {\cheng $b+1$} vertices, namely ${\cheng \hat{v}}, v_2,v_3,...,v_b$, are from $\overline{\Gamma}(\hat{v},C)$, and the others are from $\Gamma(\hat{v},C)$ in any order. Similarly, we only need to keep the first $a+1$ branches, since branch $B_{a+2}$ would have $\Delta(S_{a+2})> \tau(\sigma(B))$. 
%

{\cheng For illustration}, consider again the example in Figure~\ref{fig:branching_case}(b). Based on pivot $v_3$ in $C$ that disconnects 5 vertices in $C$, i.e., $\{v_3,v_6,v_7,v_8,v_9\}$, we define the ordering $\langle v_3,v_6,v_7,v_8,v_9,v_4,v_5 \rangle$. The branches $B_5, ..., B_8$ can be pruned since $B_5$ has $S_5=\{v_1,v_2,v_3,v_6,v_7,v_8\}$ and $\Delta(S_{5})  = 4 > \tau(\sigma(B)) = 3$.

\smallskip
\noindent\textbf{Pivot Selection.} There could be multiple vertices in $S\cup C$ with more than $\tau(\sigma(B))$ disconnections, which {\cheng are} qualified to be a pivot. 
{
We select from them the one with the largest number of disconnections {\cheng within $S\cup C$}, i.e., $\overline{\delta}(\hat{v}, S\cup C)$.
%
{\chengB We explain this strategy as follows.} 
{\chengB First}, we prune $(|C|+1) - (a+1)$ branches, i.e., $B_{a+2}, B_{a+3},...,B_{|C|+1}$.
{\chengB Second, we} also prune $b - (a+1)$ vertices from $C_{a+1}$, {\chengB i.e., $v_{a+2},v_{a+3},...,v_b$, via Refinement Rule 1}
(since including each of them, {\chengB says $v$,} to $S_{a+1}$ would have $\Delta(S_{a+1}\cup\{v\})\geq \overline{\delta}(\hat{v},S_{a+1}\cup\{v\})=\overline{\delta}(\hat{v},S)+\overline{\delta}(\hat{v},\{v_1,...,v_{a},v\})=\overline{\delta}(\hat{v},S)+(a+1)=\tau(\sigma(B))+1> \tau(\sigma(B_{a+1}))$). 
{\chengB In summary,} we would prune {\chengB more branches/vertices} when a vertex has a \emph{smaller} $a$ and/or a \emph{larger} $b$. 
{\chengB Considering $(b-a)=\overline{\delta}(\hat{v},S\cup C)-\tau(\sigma(B))$, we select the vertex with the largest $\overline{\delta}(\hat{v}, S\cup C)$ as the pivot.}
%
}
Besides, there exists at least {\cheng one} vertex in $S\cup C$ that has more than $\tau(\sigma(B))$ disconnections (i.e., $\Delta(S\cup C)>\tau(\sigma(B))$) since otherwise the branch holds a QC $G[S\cup C]$ and {\cheng the branching process} can be terminated.


{\cheng
\subsection{A Hybrid Branching Strategy: Hybrid-SE Branching}
\label{subsec:hybrid-SE}

With Sym-SE branching, we can prune those branches that violate the necessary condition (i.e., they hold no QCs). We observe that for those branches that satisfy the necessary condition (i.e., they may hold QCs), some of them may hold \emph{non-maximal} QCs only and thus they can also be pruned. Such a branch would be formed especially after \emph{excluding} a certain set of vertices from the candidate set. 
%
Specifically, we consider a scenario, where $B = (S, C, D)$ is a branch and $\hat{v}$ is a vertex in $C$ such that $\overline{\delta}(\hat{v}, S\cup C) > \tau(\sigma(B))$ and $\overline{\delta}(\hat{v}, S)=0$. We have the following lemma.
\begin{lemma}
    \label{lemma:pivse}
For any QC $G[H]$ to be found in $B$ that excludes all vertices in $\overline{\Gamma}(\hat{v},C)$ (i.e., $\forall u\in \overline{\Gamma}(\hat{v},C), u\notin H$), QC $G[H]$ is not maximal.
\end{lemma}

We consider applying SE branching and Sym-SE branching separately by selecting $\hat{v}$ as the pivot and using the ordering of vertices as specified in Equation~(\ref{sym-pivse-case2}).

\smallskip
\noindent\textbf{SE Branching.} Based on Equation~(\ref{eq:SE_branching}), we would create $|C|$ branches, which we denote by $\tilde{B}_1, \tilde{B}_2, ..., \tilde{B}_{|C|}$. We have three observations. First, branch $\tilde{B}_1$ \emph{includes} vertex $\hat{v}$ and all other branches \emph{exclude} vertex $\hat{v}$. Second, branch $\tilde{B}_{b+1}$ \emph{excludes} $D\cup \overline{\Gamma}(\hat{v}, C)$ (recall that $\overline{\Gamma}(\hat{v}, C) = \{\hat{v}, v_2, v_3, ..., v_{b}\}$), and hence branch $\tilde{B}_{b+1}$ holds no maximal QCs based on Lemma~\ref{lemma:pivse}. Therefore, we can prune $\tilde{B}_{b+1}$. Third, all branches following $\tilde{B}_{b+1}$, namely, $\tilde{B}_{b+2}, \tilde{B}_{b+3}, ..., \tilde{B}_{|C|}$, have their exclusion sets as supersets of that of branch $\tilde{B}_{b+1}$, and hence they can be pruned as well. We summarize the findings as below (branches marked in grey are those that can be pruned).
\begin{equation}
\label{eq:se-findings}
\text{SE Branching:}
\left\{
\begin{array}{ll}
      \tilde{B}_1 & \text{including $\hat{v}$} \\
      \tilde{B}_2, ..., \tilde{B}_b, {\color{lightgray}\tilde{B}_{b+1}, ..., \tilde{B}_{|C|}} & \text{excluding $\hat{v}$} \\
\end{array} 
\right. 
\end{equation}

For illustration, consider the example in Figure~\ref{fig:branching_case}(c). {\cheng The branching is based on the} pivot $v_3$, which disconnects 5 vertices in $C$, i.e., $\{v_3,v_6,v_7,v_8,v_9\}$, {\cheng and } the ordering $\langle v_3,v_6,v_7,v_8,v_9,v_4,v_5 \rangle$. The branches $B_6$ and $B_7$ can be pruned since they both exclude all vertices in $\overline{\Gamma}(v_3,C)=\{v_3,v_6,v_7,v_8,v_9\}$.

\smallskip
\noindent\textbf{Sym-SE Branching.} Based on Equation~(\ref{eq:sym-se}), we would create $|C|+1$ branches, which we denote by $\ddot{B}_1, \ddot{B}_2, ..., \ddot{B}_{|C|+1}$. Similarly, we know that branch $\ddot{B}_1$ \emph{excludes} vertex $\hat{v}$ and all other branches \emph{include} vertex $\hat{v}$. Recall that we can prune branch $\ddot{B}_{a+2}$ and all branches following $\ddot{B}_{a+2}$. We provide a summary as follows (branches marked in grey are those that can be pruned).
\begin{equation}
\label{eq:sym-se-findings}
\text{Sym-SE Branching:}
\left\{
\begin{array}{ll}
      \ddot{B}_1 & \text{excluding $\hat{v}$} \\
      \ddot{B}_2, ..., \ddot{B}_{a+1}, {\color{lightgray}\ddot{B}_{a+2}, ..., \ddot{B}_{|C|+1}} & \text{including $\hat{v}$} \nonumber \\
\end{array} 
\right. 
\end{equation}

\smallskip
\noindent\textbf{Hybrid-SE Branching.} As can be noticed, with SE branching, we can prune some branches (namely, $\tilde{B}_{b+1}, ..., \tilde{B}_{|C|}$) that hold no maximal QCs. With Sym-SE branching, we can prune some branches (namely, $\ddot{B}_{a+2}, ..., \ddot{B}_{|C|+1}$) that hold no QCs. We therefore propose a hybrid branching method based on SE branching and Sym-SE branching so as to inherit the merits of both strategies. We call this hybrid branching method \emph{Hybrid-SE branching}. Specifically, for the case of excluding the vertex $\hat{v}$, we take the branches 
$\tilde{B}_2, ..., \tilde{B}_b$ created by SE branching, and for the other case of including the vertex $\hat{v}$, we take the branches $\ddot{B}_2, ..., \ddot{B}_{a+1}$ created by Sym-SE branching. Clearly, all branches that have been taken cover all possible vertex sets under branch $B$. We provide a summary of {\revision Hybrid-SE branching} as follows (branches marked in grey are those that can be pruned).
\begin{equation}
\label{eq:hybrid-se-findings}
\text{Hybrid-SE Branching:}
\left\{
\begin{array}{ll}
      \tilde{B}_2, ..., \tilde{B}_b, {\color{lightgray}\tilde{B}_{b+1}, ..., \tilde{B}_{|C|}} &\text{excluding $\hat{v}$} \\
      \ddot{B}_2, ..., \ddot{B}_{a+1}, {\color{lightgray}\ddot{B}_{a+2}, ..., \ddot{B}_{|C|+1}} &\text{including $\hat{v}$} \\
\end{array} 
\right. 
\end{equation}
}

For illustration, consider the example in Figure~\ref{fig:branching_case}(d). {\cheng The branching takes branches $\tilde{B}_2,..., \tilde{B}_7$ created by SE branching (which exclude $v_3$) and branches $\ddot{B}_2,..., \ddot{B}_8$ created by Sym-SE branching (which include $v_3$). }
The branches $\tilde{B}_6, \tilde{B}_7$ and $\ddot{B}_5,..., \ddot{B}_8$ {\cheng can be pruned}.

{\cheng
\smallskip
\noindent\textbf{Remark.} The Hybrid-SE branching is applicable only in the case where we can find a pivot $\hat{v}$ such that $\forall u\in \overline{\Gamma}(\hat{v},C), u\notin H$ (note that we apply Hybrid-SE branching when $b=a+1$ or $\tau(\sigma(B))=1$ is also satisfied for obtaining better time complexity); otherwise, we would use the Sym-SE branching method, which is always applicable. In Section~\ref{subsec:summary_fastqc}, we will show that a BB algorithm based on our new Hybrid-SE branching (if possible) and Sym-SE branching (otherwise) would achieve new state-of-the-art worst-case time complexity. Furthermore, in the case we always use Sym-SE branching, the worst-case time complexity would be slightly worse than when we use Hybrid-SE branching if possible, but still better than when we use the existing SE branching (details can be found in the 
\ifx \CR\undefined
appendix 
\else
technical report~\cite{TR} 
\fi
for the sake of space).
}

\begin{algorithm}[t]
\small
\caption{{\cheng A new branch-and-bound algorithm}: \texttt{FastQC}}
\label{alg:FastQC}
\KwIn{A graph $G=(V,E)$, $0.5 \leq \gamma\leq 1$, and $\theta>0$}
\KwOut{{\chengrr A set of QCs that includes} all MQCs}
\texttt{FastQC-Rec}$(\emptyset,V,\emptyset)$;\\
\SetKwBlock{Enum}{Procedure \texttt{FastQC-Rec}$(S,C,D)$}{}
\Enum{
    \tcc{Progressively refining\&re-checking (Sec.~\ref{subsec:progressively_RR})}
    \Repeat{no vertices can be removed from the candidate set $C$}{
        \If{$\Delta(S)>\tau(\sigma(B))$ (Condition C1\&C2 is not satisfied)}{\textbf{return} false\;}
        Refine $B$ via Refinement Rule (1) and (2)\;
    }
    
    \tcc{Termination condition based on $\tau(\sigma(B))$ (\textbf{T1})}
    \If{$\Delta(S\cup C)\leq \tau(\sigma(B))$}{
        \lIf{{\cheng the necessary condition of maximality} is satisfied and $|S\cup C|\geq \theta$}{
            \textbf{Output} $G[S\cup C]$ 
            }
        \texttt{return} true\;
    }

    \tcc{Termination condition based on $\theta$ (\textbf{T2})}
    \lIf{any of termination conditions is satisfied}{\textbf{return} false}
    
    \tcc{Sym-SE \& Hybrid-SE branching (Sec.~\ref{subsec:Sym-PivSE} \& \ref{subsec:hybrid-SE})}
    Select a pivot $\hat{v}$ from $S\cup C$ for branching\;
    \If{$\hat{v} \in C$ and $\overline{\delta}(\hat{v},S)=0$ and ($b=a+1$ or $\tau(\sigma(B))=1$) (Hybrid-SE branching)}{
        Create $\{\ddot{B}_2,...,\ddot{B}_{a+1}, \tilde{B}_2,...,\tilde{B}_b\}$ based on Equation~(\ref{eq:hybrid-se-findings})
    }\ElseIf{$\hat{v} \in S$ (Sym-SE branching: Case 1)}{
        Create branches $\{\!\ddot{B}_1,\!\ddot{B}_2,...,\!\ddot{B}_{a+1}\!\}$ based on Equation~(\ref{eq:sym-se},\ref{sym-pivse-case1})
    }\Else{
        \tcc{Sym-SE branching: Case 2}
        Create branches $\{\!\ddot{B}_1,\!\ddot{B}_2,...,\!\ddot{B}_{a+1}\!\}$ based on Equation~(\ref{eq:sym-se},\ref{sym-pivse-case2})
    }
    \For{each branch $B_i$}{
        $\mathcal{T}_i\leftarrow$ \texttt{FastQC-Rec}$(S_i,C_i,D_i)$\;
    }

    \tcc{Additional step: output $G[S]$ if necessary}
    \If{all of $\mathcal{T}_i$ are false}{
        \If{$G[S]$ is a QC and satisfies {\cheng the necessary condition of} maximality}{
            \textbf{Output} $G[S]$ if $|S|\geq \theta$; \textbf{return} true;   
        }
        \textbf{return} false;
    }
    \textbf{return} true;
}

\end{algorithm}

\subsection{\texttt{FastQC}: Summary and Analysis}
\label{subsec:summary_fastqc}
{\cheng Based on the newly proposed pruning techniques (namely the the progressive procedure of refining a branch and re-checking the necessary condition) and branching methods (namely Hybrid-SE and Sym-SE), we develop a new BB algorithm called \texttt{FastQC}.}
The pseudocode of \texttt{FastQC} is presented in Algorithm~\ref{alg:FastQC}, which differs from \texttt{Quick+} in the following aspects. \underline{First}, it progressively refines a branch and re-checks the necessary condition until a refined branch is pruned or a branch cannot be refined any further (line 3-7). \underline{Second}, if a refined branch satisfies the necessary condition and {\cheng is not} pruned, we then {\cheng check two termination conditions, namely T1} based on the obtained $\tau(\sigma(B))$ and {\cheng T2 based on the} size constraint $\theta$ {\cheng and terminate the branch if any of the condition is satisfied} (line 8-11). 

\smallskip
\noindent\textbf{T1: Termination condition based on $\tau(\sigma(B))$.} As discussed earlier, we can terminate the branch when $\Delta(S\cup C)\leq \tau(\sigma(B))$ (line 8-10) since the branch holds a QC $G[S\cup C]$ and any other QC under this branch is a subgraph of $G[S\cup C]$. 
In this case, 
{\cheng we check the following necessary condition for a QC $G[H]$ to be maximal,}
\begin{equation}
   \nexists\  v\in V-H,\ G[H\cup v]\ \text{is a QC},\nonumber
\end{equation}
which can be done in polynomial time (note that the problem of checking the maximality of a QC exactly is NP-hard~\cite{sanei2018enumerating}).
{\cheng If yes and the size of $G[S\cup C]$ is at least $\theta$, we output $G[S\cup C]$.}

\smallskip
\noindent\textbf{T2: Termination condition based on size constraint $\theta$.} During the recursive procedure, if any of the following two conditions is satisfied, we can terminate the branch (line 11) since no MQC with size at least $\theta$ would be found.
\begin{enumerate}[leftmargin=*]
    \item $|S\cup C|< \theta$.
    \item There exists a vertex $v\in S$ such that $\delta(v,S\cup C)<\theta-\tau(\sigma(B))$.
\end{enumerate}
For simplicity, we put the proof in the 
\ifx \CR\undefined
appendix.
\else
technical report~\cite{TR}.
\fi

\underline{Third}, we select a pivot from $S\cup C$ and conduct the Hybrid-SE branching (if possible) {\cheng and} Sym-SE branching {\cheng (otherwise)} for forming sub-branches (line 12-20). 

Note that \texttt{FastQC} would also need to monitor whether a sub-branch of the current one would find a QC (line 21-25), similarly as \texttt{Quick+} {\cheng does}. In addition, \texttt{FastQC} would return a superset of all MQCs which inevitably contains some non-maximal QCs, 
{\revision {\chengr i.e., \texttt{FastQC} solves the MQCE-S1 problem, but not the MQCE problem, as \texttt{Quick+} does.}}



\smallskip
\noindent\textbf{Worst-case time complexity.} The worst-case time {\cheng complexity} of \texttt{FastQC} is strictly smaller than that of \texttt{Quick+}. We give the details in the following theorem (with the proof provided in the 
\ifx \CR\undefined
appendix for simplicity).
\else
technical report~\cite{TR}).
\fi
\begin{theorem}
\label{theorem:fastqc}
    Given a graph $G=(V,E)$, \texttt{FastQC} finds {\chengrr a set of QCs that includes} all MQCs with the size at least $\theta$, {\revision {\chengr i.e., it solves the MQCE-S1 problem,}} in $O(n\cdot d \cdot \alpha_{k}^{n})$ time where $\alpha_k$ is the largest real root of $x^{k+2}-x^{k+1}-2x^k+2=0$ (when $k\geq 2$) and $k=\tau(n)$ is an upper bound of the largest $\tau(\sigma(B))$ (i.e., $\tau(\sigma(B))\leq k$ for any branch $B$). {\cheng $\alpha_{k}$ is strictly smaller than 2.} For example, when $k=$ 2, 3 and 4, $\alpha_{k}=$ 1.769, 1.899, and 1.953 respectively. Besides, when $k=1$, $\alpha_{k}=1.445$.
\end{theorem}

{\cheng
\smallskip
\noindent\textbf{Remark 1.}
We note that there exist some studies, which break the $O^*(2^{n})$ worst-case time complexity for enumerating subgraphs that satisfy the hereditary property~\cite{zhou2020enumerating,yu2022maximum}. We emphasize that (1) these methods cannot be directly applied to our problem of enumerating QCs which do not satisfy the hereditary property; (2) our method is the first one which breaks the $O^*(2^{n})$ worst-case time complexity for enumerating QCs; and (3) the constants $\alpha_k$ in the time complexity of our method (e.g., our smallest constant is 1.769 when $k=2$) are smaller than those of many existing methods~\cite{zhou2020enumerating,yu2022maximum} (e.g., their smallest constant is 1.839 for $k=2$), which is due to the newly proposed necessary condition in the SD space and the branching methods.

{\revision
\smallskip
\noindent\textbf{Remark 2.} 
We remark that {\chengr for solving the MQCE problem,} the worst-case time complexity 
is $O^*(\alpha_k^{n}+\min\{|\mathcal{S}_{fast}|^2,~~|\mathcal{S}_{fast}|\cdot 2^{2\omega}\})$ (resp. $O^*(2^{n}+\min\{|\mathcal{S}_{quick}|^2,~~|\mathcal{S}_{quick}|\cdot 2^{2\omega}\})$) when adopting \texttt{FastQC} (resp. \texttt{Quick+}) {\chengr for solving MQCE-S1 and the method in~\cite{savnik2021data} for sovling MQCE-S2},
where $\mathcal{S}_{fast}$ (resp. $\mathcal{S}_{quick}$) is the set of QCs returned by \texttt{FastQC} (resp. \texttt{Quick+}). 
We note that $|\mathcal{S}_{fast}|$ and $|\mathcal{S}_{quick}|$ can be bounded by the number of branches produced by \texttt{FastQC} and \texttt{Quick+}, i.e., $O^*(\alpha_k^{n})$ and $O^*(2^{n})$, respectively, since {\chengrr at most one QC can be returned per branch}.
{\chengr Since $\omega$ is bounded by $n$, we deduce that the time complexity of solving MQCE with \texttt{FastQC} is $O^*(\alpha_k^{2n})$ and that with \texttt{Quick+} is $O^*(2^{2n})$. Furthermore, for sparse graphs with $\omega$ bounded by $O(\log n^c)$ where $c$ is a constant, the time complexity of solving MQCE with \texttt{FastQC} is $O^*(\alpha_k^{n})$ and that with \texttt{Quick+} is $O^*(2^{n})$. In any case, the method based on \texttt{FastQC} has a strictly smaller theoretical time complexity than that based on \texttt{Quick+}. Based on our experimental results, the former runs faster than the latter by up to two orders of magnitude.}
}
}
\section{A Divide-and-Conquer Framework with \texttt{FastQC}: \texttt{DCFastQC}}
\label{sec:DC_Framework}

{\cheng While \texttt{FastQC} has a lower time complexity than existing methods (e.g., \texttt{Quick+}), it may still suffer from a scalability issue when running on big graphs. To further boost the efficiency and scalability of finding MQCs, we adopt a \emph{divide-and-conquer} strategy, which is to divide the whole graph into multiple smaller ones and then run \texttt{FastQC} on each of them. We call the resulting algorithm \texttt{DCFastQC}, which guarantees to find all MQCs. Furthermore, we develop some pruning techniques to shrink the constructed smaller graphs for better efficiency. In summary, \texttt{DCFastQC} would invoke \texttt{FastQC} multiple times, each on a smaller graph (compared with the original graph), and thus the scalability is improved. We note that this \emph{divide-and-conquer} strategy has been widely used for enumerating subgraphs~\cite{guo2020scalable, yu2022maximum, zhou2020enumerating, khalil2022parallel}. Our technique differs from existing ones in (1) the way of how a graph is divided~\cite{guo2020scalable,khalil2022parallel}; and/or (2) the techniques for shrinking the smaller graphs~\cite{guo2020scalable, yu2022maximum, zhou2020enumerating, khalil2022parallel}.}

To be specific, given an ordering ${\cheng \langle} v_1,v_2,...,v_{|V|} {\cheng \rangle}$, it divides the whole graph $G$ into $|V|$ subgraphs, namely $G_i=G[V_i]$ for $1\leq i\leq |V|$, as follows.
\begin{equation}
    \label{eq:decomposition}
    V_i=\Gamma_2(v_i,V)-\{v_1,v_2,...,v_{i-1}\},
\end{equation}
where $\Gamma_2(v_i,V)$ is the set of 2-hop neighbours of $v_i$ in $V$ and $|V_i|$ is thus bounded by $O(d^2)$. 
{\cheng Then, on} each subgraph $G_i$, {\cheng it runs} \texttt{FastQC} by starting with the {\cheng branch $B = (S, C, D)$} with $S=\{v_i\}$, $C=V_i-\{v_i\}$ and $D=\{v_1,v_2,...,v_{i-1}\}$. {\cheng Note that} all MQCs found in $G_i$ would include $v_i$ and exclude $\{v_1,v_2,...,v_{i-1}\}$. It is {\cheng not difficult} to verify that {\cheng each} MQC would be found exactly once from one of above subgraphs based on Property~\ref{property:diameter} (for which we put the proof in the 
\ifx \CR\undefined
appendix 
\else
technical report~\cite{TR} 
\fi
{\cheng for the sake of space}).

{\cheng The framework} can be further improved by {\cheng shrinking the} subgraphs formed as above, {\cheng with techniques of} \emph{vertex ordering} and \emph{pruning rules on} $G_i$ as {\cheng presented} below.

\smallskip
\noindent\textbf{Degeneracy ordering.} 
{\cheng By following some existing studies~\cite{zhou2020enumerating, yu2022maximum}, we adopt the degeneracy ordering of $V$ for dividing a graph.}
The reason is two-fold. First, the size of each subgraph $|V_i|$ {\cheng would be} bounded by $O(\omega  d)$ based on the property of degeneracy ordering where $\omega$ denotes the degeneracy of $G$~\cite{zhou2020enumerating, yu2022maximum}. Second, the degeneracy ordering can be obtained by core decomposition in polynomial time $O(|E|)$ {\cheng efficiently}~\cite{batagelj2003m}.

\smallskip
\noindent\textbf{Pruning rules on $G_i$.} We can prune the following vertices from a subgraph $G_i$.
\begin{itemize}[leftmargin=*]
    \item {\cheng \textbf{One-hop pruning.}} $u\in V_i-\{v_i\}$, $\delta(u,V_i)<\lceil \gamma\cdot (\theta-1) \rceil$.
    \item {\cheng \textbf{Two-hop pruning.}} $u\in V_i-\{v_i\}$, (1) if $u\in \Gamma(v_i,V_i)$, $|\Gamma(v_i,V_i)\cap \Gamma(u,V_i)|< f(\theta)$ or (2) if $u\notin \Gamma(v_i,V_i)$, $|\Gamma(v_i,V_i)\cap \Gamma(u,V_i)|< f(\theta)+2$, where
    $f(\theta)=\theta-\tau(\theta) - \tau(\theta+1)$.
\end{itemize}
 We put the proof of above pruning rules in the 
 \ifx \CR\undefined
 appendix 
 \else
 technical report~\cite{TR} 
 \fi
 {\cheng for the sake of space}. 
 Moreover, we can iteratively apply one-hop pruning and two-hop pruning on $G_i$ for multiple rounds, which would boost their effectiveness. The rationale is that with some vertices excluded from $G_i$ in a former round, the degrees of the remaining vertices would become smaller and thus {\cheng they can potentially be pruned in the current round.}
 
\begin{algorithm}{}
\small
\caption{{\cheng A divide-and-conquer framework with \texttt{FastQC}}: \texttt{DCFastQC}}
\label{alg:DCFastQC}
\KwIn{A graph $G=(V,E)$, $0.5 \leq \gamma\leq 1$, and $\theta>0$}
\KwOut{{\chengrr A set of QCs that includes} all MQCs}
{\cheng Reduce} $G=(V,E)$ as a $\lceil \gamma \cdot (\theta-1) \rceil$-core of $G$\;
{\cheng Compute the degeneracy ordering $\langle v_1,v_2,...,v_n \rangle$\;}
\For{ each $v_i$ in $\{v_1,v_2,...,v_n\}$}{
    Construct $G_i=G[V_i]$ based on Equation~(\ref{eq:decomposition})\;
    \For{{\cheng $i = $} 1, 2, ..., MAX\_ROUND}{
        Refine $V_i$ by one-hop {\cheng pruning} and two-hop pruning;
    }
    Construct $S=\{v_i\}$, $C=V_i-\{v_i\}$ and $D=\{v_1,...,v_{i-1}\}$\;
    FastQC-Rec$(S,C,D)$\;
}
\end{algorithm}

{\cheng
\smallskip
\noindent\textbf{The \texttt{DCFastQC} Algorithm.}
The pseudocode of \texttt{DCFastQC} is presented in Algorithm~\ref{alg:DCFastQC}. First, it reduces the graph to be the $\lceil \gamma\cdot (\theta - 1) \rceil$-core of $G$ (line 1). This is because every QC with size at least $\theta$ is within the $\lceil \gamma\cdot (\theta - 1) \rceil$-core of $G$~\cite{guo2020scalable}. Then, it computes the degeneracy ordering (line 2). Finally, it performs $n$ iterations (line 3 - 8). At the $i^{th}$ iteration, it constructs a smaller graph $G_i = G[V_i]$ (line 4), prunes the vertices from $V_i$ for MAX\_ROUND rounds, where MAX\_ROUND is a user parameter for controlling the trade-off between the workload and the effectiveness of the pruning techniques (line 5-6), and then runs \texttt{FastQC} on the refined graph (line 7-8).
}

\smallskip
\noindent\textbf{Time complexity.} 
The time cost is dominated by the part of invoking \texttt{FastQC} {\cheng $O(n)$ times}. 
{\cheng Recall that the number of vertices in a graph $G_i$ is} 
bounded by $O(\omega  d)$ {\cheng (as analyzed earlier)}. 
{\cheng Based on the time complexity of \texttt{FastQC} presented in Theorem~\ref{theorem:fastqc}, we deduce that} the time complexity of \texttt{DCFastQC} is $O(n\cdot \omega d^2 \cdot \alpha_{k}^{\omega  d})$ where $\alpha_k$ is the largest real root of $x^{k+2}-x^{k+1}-2x^k+2=0$ (when $k\geq 2$) and $k= \lfloor \omega(1-\gamma)/\gamma +1 \rfloor$ (the proof is put in the 
\ifx \CR\undefined
appendix for simplicity).
\else
technical report~\cite{TR} for simplicity).
\fi
%
%
%
%
%
We remark that {\cheng in practice}, \texttt{DCFastQC} is faster than \texttt{FastQC} since large graphs usually have $\omega$ and $d$ far smaller than the total number of vertices, {\cheng which will be verified in our experiments}.



\section{Experimental Results}
\label{sec:exp}

\begin{table*}[h]
	\centering
	\scriptsize
	\small
	\caption {Real datasets}
	\label{tab:dataset}
 	\vspace{-0.10in}
	\begin{tabular}{c|r|r|c|c|c|c|c|c|c|c|c|c|c}
		\hline
		 \multicolumn{1}{c|}{{Dataset}}
		 & \multicolumn{1}{c|}{$|V|$}
		 & \multicolumn{1}{c|}{$|E|$}
		 & \multicolumn{1}{c|}{$|E|/|V|$}
		 & \multicolumn{1}{c|}{$d$}
          & \multicolumn{1}{c|}{$\ \omega\ \ $}
		 & \multicolumn{1}{c|}{$\ \ \theta_{d}\ \ $}
		 & \multicolumn{1}{c|}{$\ \ \gamma_{d}\ \ $}
          & \multicolumn{1}{c|}{\#\{MQC\}}
          & \multicolumn{1}{c|}{\#\{DCFastQC\}}
          & \multicolumn{1}{c|}{\#\{Quick+\}}
          & \multicolumn{1}{c|}{\revision $|H_{min}|$}
          & \multicolumn{1}{c|}{\revision $|H_{max}|$}
          & \multicolumn{1}{c}{\revision $|H_{avg}|$}
		\\
		\hline\hline
		Ca-GrQC   & 5,242  & 14,496  & 2.77 &  81 & 43 &  10 &  0.9 &  1,665 & 1,725 & 2,232 & \revision 10	& \revision 46 & \revision 26.56 \\
		\hline
		Opsahl   & 2,939  & 15,677 & 5.33 & 473 & 28 & 20 & 0.9 & 34,508 &35,681 & 263,943 & \revision 21	& \revision 26 & \revision 21.69\\
		\hline
		CondMat  & 39,577  & 175,691  & 4.43 &278 & 29 & 10 & 0.9 & 7,222 &  7,977 & 11,465 & \revision 10	& \revision 30 & \revision 13.33\\
		\hline
		\textbf{Enron}  & 36,692  & 183,831  & 5.01 &1383 & 43 & 23 & 0.9 &200 & 212 & 335 & \revision 23	& \revision 24 & \revision 23.08\\
		\hline
		Douban   & 154,908  & 327,162 & 2.11 &287 & 15 &12 &0.9 &26 & 26 & 26 & \revision 12	& \revision 12 & \revision 12\\
		\hline
		\textbf{WordNet}  & 146,005  & 656,999  & 4.49 &1008 & 31 & 14 &0.9 &2,515  & 2,691 & 5,231 & \revision 14	& \revision 32 & \revision 17.29\\
		\hline
		Twitter  & 465,017  & 833,540  & 1.79 &677 & 30 &6 &0.9 &11 & 11 & 11 & \revision 6	& \revision 6 & \revision 6\\
		\hline
		\textbf{Hyves}    & 1,402,673  & 2,777,419  & 1.98 &31,883& 39 &23 &0.9 &114 & 117 & 168 & \revision 23	& \revision 24 & \revision 23.05\\
		\hline
		Trec  & 1,601,787  & 6,679,248  & 1.98 &25,609& 140 & 50 &0.96 &682,736  & 682,862 & 2,659,161 & \revision 51	& \revision 91 & \revision 54.64\\
		\hline
		Flixster  & 2,523,386  & 7,918,801 & 3.14 & 1,474 & 123& 35 & 0.96 & 22,853 & 24,829 & 52,845 & \revision 35	& \revision 38 & \revision 35.16\\
		\hline
		\textbf{Pokec}   & 1,632,803  & 22,301,964  & 13.66 & 20,518 & 47 & 32 & 0.9 & 7 & 7 & 7 & \revision 32	& \revision 32 & \revision 32\\
		\hline
	    FullUSA  & 23,947,347  & 28,854,312 & 1.20 & 9& 3 & 3 & 0.51 & 35   & 35 & 35 & \revision 6	& \revision 6 & \revision 6\\
		\hline
        \revision Kmer   & \revision 67,716,231  & \revision 69,389,281 & \revision 1.02 & \revision 35& \revision 6 & \revision 10 & \revision 0.51 & \revision 146   & \revision 176 & \revision 265 & \revision 10	& \revision 12 & \revision 10.09\\
        \hline
        \revision UK2002   & \revision 18,483,186  & \revision 261,787,258 & \revision 14.16 & \revision 194,955& \revision 943 & \revision 450 & \revision 0.96 & \revision 6   & \revision 27 & \revision --- & \revision 475	& \revision 944 & \revision 651\\
        \hline
	\end{tabular}
\end{table*}

\noindent\textbf{Datasets.} We {\cheng use} both real and synthetic datasets in experiments. The real datasets are {\cheng collected from http://konect.cc/ and come from} different domains. {\cheng The statistics of the real datasets are} summarized in Table~\ref{tab:dataset}, where the edge density of a graph $G=(V,E)$ is defined by $|E|/|V|$, $d$ denotes the maximum degree, $\omega$ represents the graph degeneracy, $\theta_d$ and $\gamma_d$ are default settings {\cheng of $\theta$ and $\gamma$, respectively}. 
%
%
The synthetic datasets are generated {\cheng based on} the Erd{\"o}s-R{\'e}yni (ER) graph model. Specifically, we first generate a certain number of vertices and then randomly add a certain number of edges between pairs of vertices. By default, the number of vertices and edge density are set as 100k and 20 for synthetic datasets, respectively.

{\revision
\smallskip
\noindent\textbf{Statistics of large MQCs.} The statistics of large MQCs in the real datasets are {\chengr provided} in Table~\ref{tab:dataset}, where \#\{MQC\} denotes the number of large $\gamma_d$-MQCs with the size at least $\theta_d$ and $|H_{min}|$, $|H_{max}|$ and $|H_{avg}|$ denote the minimum, maximum and average size of MQCs in the datasets, respectively. 
We remark that the number of \emph{large} MQCs would decrease significantly as $\theta$ grows (details {\chengr can be found} in the 
\ifx \CR\undefined
appendix) 
\else
technical report~\cite{TR}) 
\fi
and thus is far smaller than the exponential {\chengr in $n$ under} our settings of $\theta_d$.  
Besides, the found MQCs are usually sufficiently large to be meaningful (with at least 10 and up to 944 vertices for the most datasets). We note that the largest MQC found in Twitter ($\gamma=0.9$) and FullUSA $(\gamma=0.51)$ contains 6 vertices since they are quite sparse and do not have any locally dense region. We use them mainly for testing the efficiency and scalability of our algorithm.
}

\smallskip
\noindent\textbf{Algorithms.}  We compare our proposed algorithm \texttt{DCFastQC} {\cheng with} \texttt{Quick+}~\cite{khalil2022parallel}. \texttt{Quick+} is the state-of-the-art algorithm as introduced in Section~\ref{sec:quick-plus}, which runs significantly faster than previous methods, {\cheng including} \texttt{Crochet}~\cite{pei2005mining,jiang2009mining}, \texttt{Cocain}~\cite{zeng2006coherent}, \texttt{Quick}~\cite{liu2008effective}. 
{\cheng We also compare different branching strategies
and different divide-and-conquer frameworks, including the one proposed in this paper and the one proposed in~\cite{khalil2022parallel,guo2020scalable}.}
{Besides, we use the set containment query algorithm proposed in~\cite{savnik2021data} for implementing the post-processing step for filtering out the non-maximal outputs.}

\smallskip
\noindent\textbf{{\cheng Implementation and Settings}.} 
All algorithms {\cheng are} implemented in C++ and tested on a Linux machine with a 2.10GHz Intel CPU and 128GB memory. We use the recent implementation of \texttt{Quick+}~\cite{khalil2022parallel}.
We measure and compare the running {\cheng times} of the algorithms under various settings. 
By following existing studies~\cite{liu2008effective,khalil2022parallel,guo2020scalable}, we report the running time that excludes the time for filtering out non-maximal QCs since it can be done efficiently~\cite{khalil2022parallel} (e.g., it can be finished within 16s for all datasets used in our experiments).
We set the running time limit (INF) as 24 hours and select four datasets, {\cheng namely Enron, WordNet, Hyves and Pokec,} as default ones {\cheng since they} cover different graph sizes and edge {\cheng densities)}. 
Besides, the default settings of parameter $\gamma$ and $\theta$ are given in Table~\ref{tab:dataset} for each dataset, which are determined based on the graph statistics.
{\cheng For example}, Trec and Flixster have a larger default value of $\gamma$ (i.e., 0.96) since the number of MQCs grows exponentially {\cheng when $\gamma$ decreases} and {\cheng there exist} a significant number of 0.96-MQCs. {\cheng In contrast}, FullUSA has {\cheng a smaller} default value of $\gamma$ (i.e., 0.51) since it is very sparse and thus has few MQCs {\cheng for} a large $\gamma$.
Our code and datasets are available at https://github.com/KaiqiangYu/SIGMOD24-MQCE.

\begin{figure}[]
	\centering
	\includegraphics[width=0.98\linewidth]{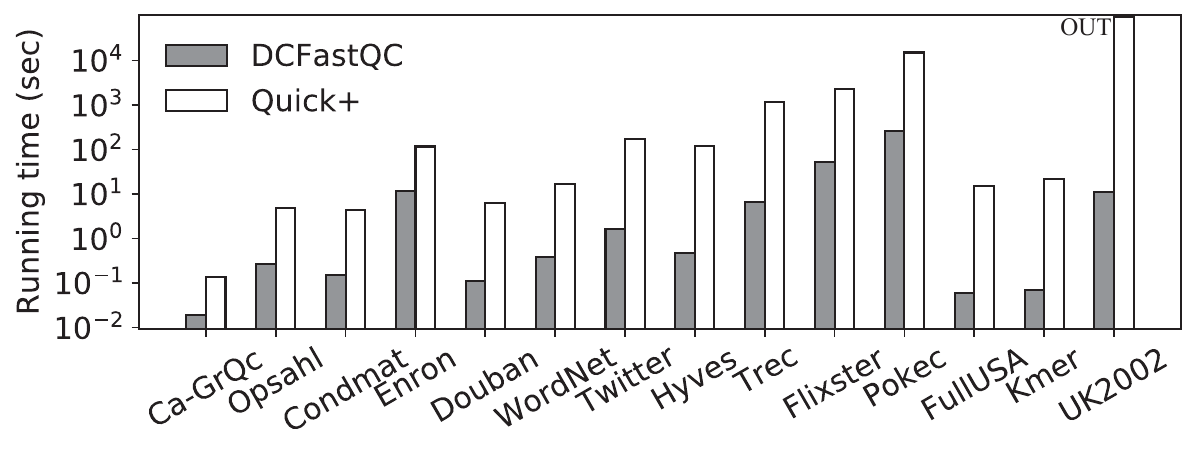}
	\vspace{-0.10in}
	\caption{\revision Comparison on all real datasets}
	\label{fig:all_datasets}
\end{figure}

\subsection{Comparison among Algorithms}
\label{subsec:CompAmongAlg}

\noindent\textbf{All datasets {\cheng (Default Settings)}.}
We compare our algorithm \texttt{DCFastQC} with the baseline \texttt{Quick+} on various datasets using default $\gamma_d$ and $\theta_d$ settings {\cheng as shown} in Table~\ref{tab:dataset}. We report the running time in Figure~\ref{fig:all_datasets} and the number of returned QCs, denoted by \#\{DCFastQC\} and \#\{Quick+\}, in Table~\ref{tab:dataset}. {\revision We observe that (1) our algorithm \texttt{DCFastQC} outperforms \texttt{Quick+} on all datasets and achieves up to {\chengr 100x} speedup and (2) \texttt{Quick+} runs out of the 128GB memory budget (denoted by OUT) and cannot finish on the largest dataset UK2002.} This observation demonstrates the efficiency and scalability of \texttt{DCFastQC} in practice and is also compatible with the theoretical results that \texttt{DCFastQC} has the worst-case running time strictly smaller than {\cheng that of} \texttt{Quick+}. 
Besides, \texttt{DCFastQC} has the number of outputs almost {\cheng the} same as that of MQCs, and outputs {\cheng fewer} non-maximal QCs compared with \texttt{Quick+}. This is mainly because the necessary condition of maximality would prune many non-maximal outputs. {\cheng For example, on the} dataset Opsahl with 34k MQCs inside, \texttt{DCFastQC} returns 35k QCs while \texttt{Quick+} returns 263k QCs. Consequently, the post-processing step of \texttt{DCFastQC} runs faster than that of \texttt{Quick+} (the results are put in the 
\ifx \CR\undefined
appendix 
\else
technical report~\cite{TR} 
\fi
for simplicity since it can be done quickly within 0.1 second on most datasets). 
{\revision Finally, we observe that \texttt{DCFastQC} would run slower on a denser graph (e.g., Enron) while running faster on a {\chengr sparser} graph (e.g., Douban). This is because the time cost of \texttt{DCFastQC} is $O(n\cdot \omega d^2 \cdot \alpha_k^{\omega d})$ and the values of $d$ and $\omega$ of a {\chengrr denser} graph tend to be larger.
}

\begin{figure}[]
	\centering
	\begin{tabular}{c c}
		\begin{minipage}{3.80cm}
			\includegraphics[width=4.1cm]{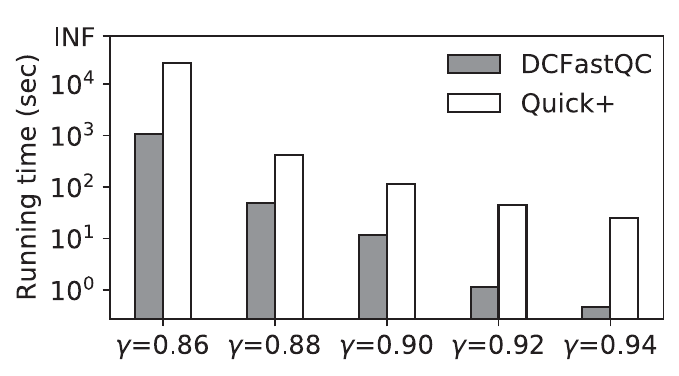}
		\end{minipage}
		&
		\begin{minipage}{3.80cm}
			\includegraphics[width=4.1cm]{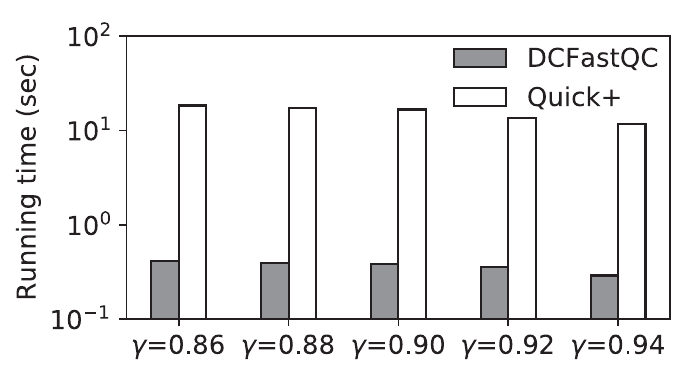}
		\end{minipage}		
		\\
		(a) Varying $\gamma$ (Enron)
		&
		(b) Varying $\gamma$ (WordNet)
		\\
		\begin{minipage}{3.80cm}
			\includegraphics[width=4.1cm]{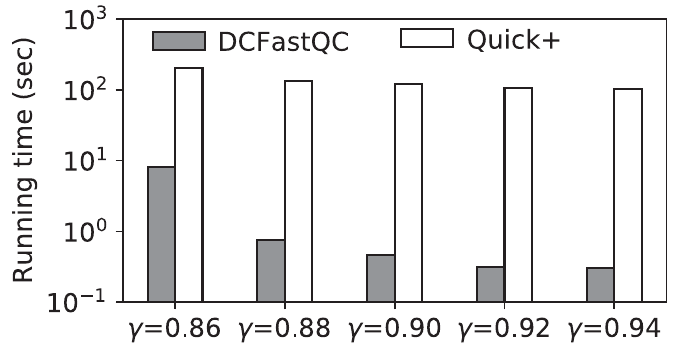}
		\end{minipage}
		&
		\begin{minipage}{3.80cm}
			\includegraphics[width=4.1cm]{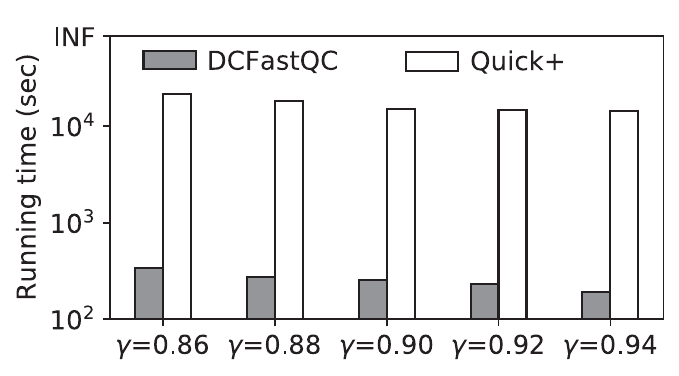}
		\end{minipage}		
		\\
		(c) Varying $\gamma$ (Hyves)
		&
		(d) Varying $\gamma$ (Pokec) 
	\end{tabular}
	\vspace{-0.15in}
	\caption{Comparison by varying $\gamma$}
	\label{fig:vary_gamma}
        \vspace{-0.05in}
\end{figure}
\smallskip
\noindent\textbf{Varying $\gamma$.} We report the running time in Figure~\ref{fig:vary_gamma} as $\gamma$ varies. We have the following observations. First, \texttt{DCFastQC} significantly outperforms \texttt{Quick+} by achieving up to two orders of magnitude speedup. Second, the running {\cheng times} of all algorithms {\cheng usually drop} as $\gamma$ {\cheng increases}. This is because the number of MQCs {\cheng decreases} exponentially {\cheng as} $\gamma$ {\cheng increases}. Third, the achieved speedup increases {\cheng as} $\gamma$ {\cheng increases}, which indicates that \texttt{DCFastQC} performs better for lager $\gamma$'s. Possible reasons include (1) the parameter $\tau(\sigma(B))$ (with the value equal to $\min\{\lfloor |S\cup C|\cdot (1-\gamma)+\gamma \rfloor, \lfloor d_{min}(B)\cdot (1-\gamma)/\gamma + 1 \rfloor\}$ ) decreases as $\gamma$ grows and correspondingly the pruning rules based on $\tau(\sigma(B))$ become more effective; (2) our branching strategy would produce fewer branches for larger $\gamma$'s according to the theoretical results, i.e., the number of formed branches in the worst case is bounded by $O^*(\alpha_k^{\omega d})$ {\cheng (details can be found in the proof of the time complexity of \texttt{FastQC}, which is put in the 
\ifx \CR\undefined
appendix) 
\else
technical report~\cite{TR}) 
\fi
}. Note that the parameter $k$ (with the value of $\min\{\lfloor\omega d (1-\gamma)+\gamma \rfloor, \lfloor \omega(1-\gamma)/\gamma +1 \rfloor\}$) decreases as $\gamma$ grows and $\alpha_k$ becomes slightly smaller.

\begin{figure}[]
	\centering
	\begin{tabular}{c c}
		\begin{minipage}{3.80cm}
			\includegraphics[width=4.1cm]{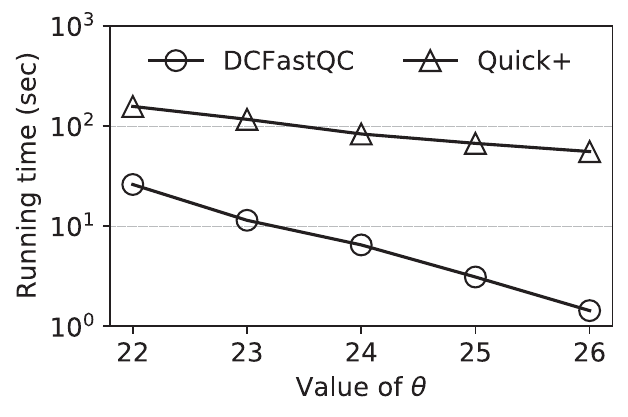}
		\end{minipage}
		&
		\begin{minipage}{3.80cm}
			\includegraphics[width=4.1cm]{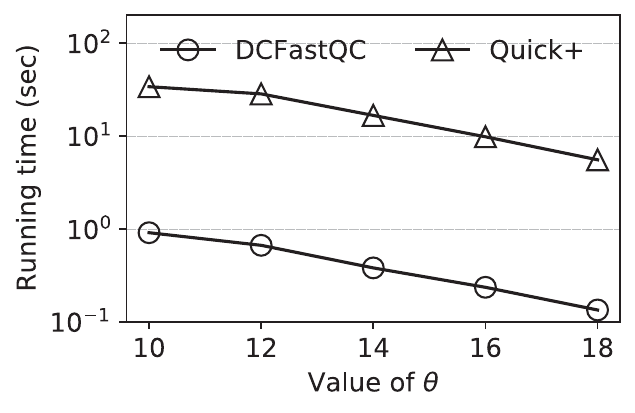}
		\end{minipage}		
		\\
		(a) Varying $\theta$ (Enron)
		&
		(b) Varying $\theta$ (WordNet)
		\\
		\begin{minipage}{3.80cm}
			\includegraphics[width=4.1cm]{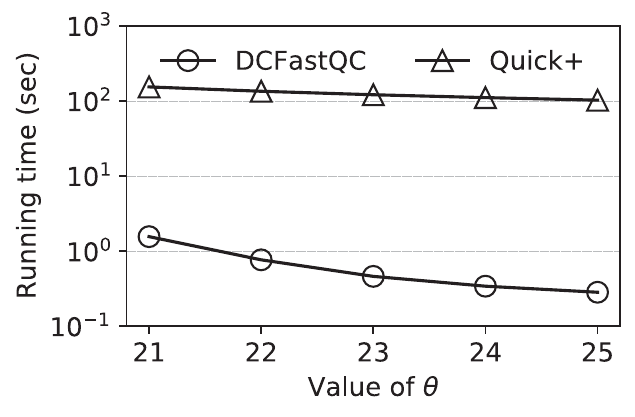}
		\end{minipage}
		&
		\begin{minipage}{3.80cm}
			\includegraphics[width=4.1cm]{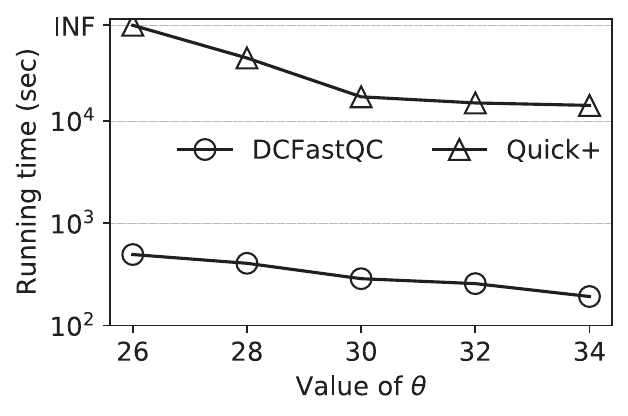}
		\end{minipage}		
		\\
		(c) Varying $\theta$ (Hyves)
		&
		(d) Varying $\theta$ (Pokec) 
	\end{tabular}
	\vspace{-0.10in}
	\caption{Comparison by varying $\theta$}
	\label{fig:vary_theta}
\end{figure}
\smallskip
\noindent\textbf{Varying size threshold $\theta$.} We report the running time in Figure~\ref{fig:vary_theta} as $\theta$ varies. Our algorithm \texttt{DCFastQC} outperforms \texttt{Quick+} by achieving up to two orders of magnitude speedup on various settings. In addition, the running {\cheng times} of all algorithms {\cheng drop} as $\theta$ {\cheng increases}. This is mainly because (1) the number of large QCs (with the size at least $\theta$) {\cheng decreases} exponentially with the {\cheng increase} of $\theta$; (2) the pruning techniques based on $\theta$ and the proposed \texttt{DC} framework are more effective for larger $\theta$'s.

\begin{figure}[]
	\centering
	\begin{tabular}{c c}
		
		\begin{minipage}{3.80cm}
			\includegraphics[width=4.1cm]{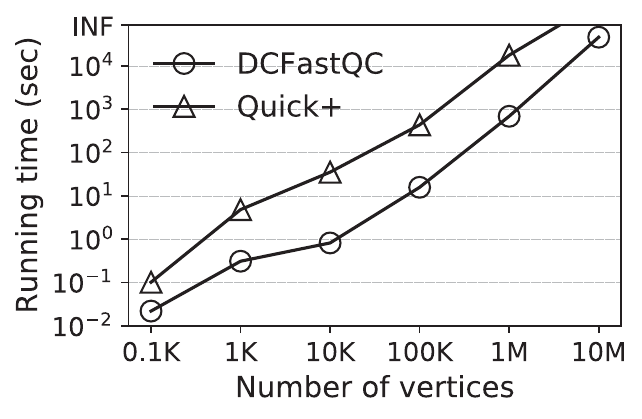}
		\end{minipage}
		&
		\begin{minipage}{3.80cm}
			\includegraphics[width=4.1cm]{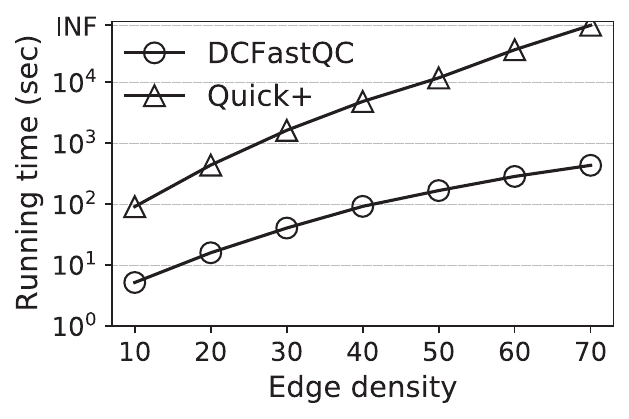}
		\end{minipage}		
		\\
		(a) Varying \# of vertices
		&
		(b) Varying edge density
	\end{tabular}
	\vspace{-0.10in}
	\caption{Comparison on synthetic datasets}
	\label{fig:synthetic}
\end{figure}
\smallskip
\noindent\textbf{Varying \# of vertices (scalability test on synthetic datasets).} We test the scalability on synthetic datasets using default settings of $\gamma_d=0.9$ and $\theta_d=10$  and report the running time in Figure~\ref{fig:synthetic}(a) as the number of vertices varies. \texttt{DCFastQC} is faster than \texttt{Quick+} by achieving at least 10$\times$ speedup and can handle the largest datasets within INF while \texttt{Quick+} cannot. In addition, the running time increases as the graph scale becomes larger. 

\smallskip
\noindent\textbf{Varying edge density.} We use default settings of $\gamma_d=0.9$ and $\theta_d=10$ and report the running time in Figure~\ref{fig:synthetic}(b) as the edge density varies. 
We have the following observations. First, \texttt{DCFastQC} runs faster than \texttt{Quick+} by achieving up to 1000$\times$ speedup and can handle the densest datasets with the edge density $|E|/|V|$ up to 70 while \texttt{Quick+} cannot. Second, the running time clearly rises as the graph becomes denser. The reason is two-fold: (1) the number of MQCs increases as the edge density grows and (2) those pruning rules based on the degree of vertices are {\cheng less} effective for {\cheng denser} graphs since the vertices have the degree increase as the graph becomes denser and thus are hard to be pruned. Third, \texttt{DCFastQC} achieves {\cheng higher} speed-ups as the graph becomes denser.

\subsection{Performance Study}

\begin{figure}[]
	\centering
	\begin{tabular}{c c}
		\begin{minipage}{3.80cm}
			\includegraphics[width=4.1cm]{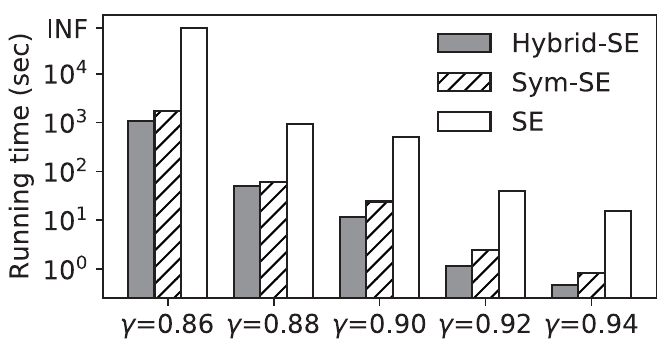}
		\end{minipage}
		&
		\begin{minipage}{3.80cm}
			\includegraphics[width=4.1cm]{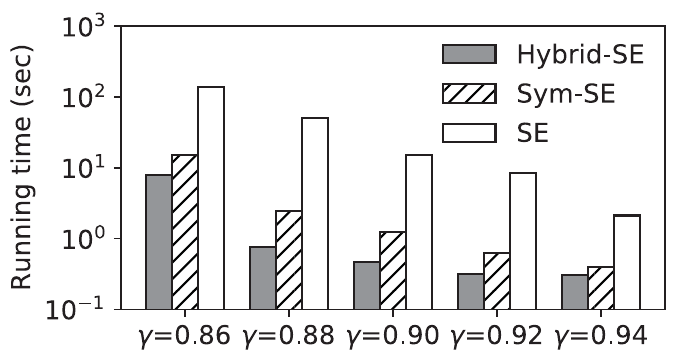}
		\end{minipage}		
		\\
		(a) Varying $\gamma$ (Enron)
		&
		(b) Varying $\gamma$ (Hyves)
		\\
		\begin{minipage}{3.80cm}
			\includegraphics[width=4.1cm]{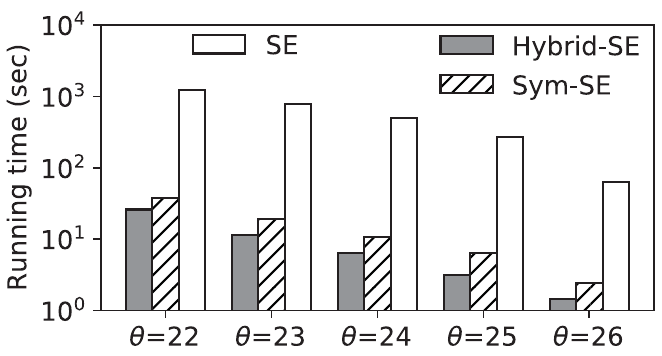}
		\end{minipage}
		&
		\begin{minipage}{3.80cm}
			\includegraphics[width=4.1cm]{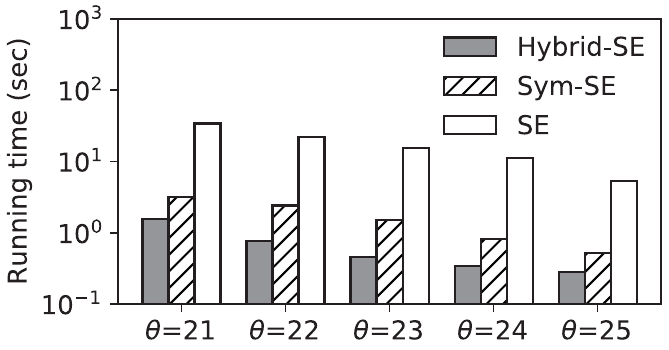}
		\end{minipage}		
		\\
		(c) Varying $\theta$ (Enron)
		&
		(d) Varying $\theta$ (Hyves) 
	\end{tabular}
	\vspace{-0.10in}
	\caption{Comparison among various branching strategies}
	\label{fig:comparsion_branching}
\end{figure}

\noindent\textbf{Comparison among various branching strategies.} We study the {\cheng effects} of various branching strategies 
{\cheng by comparing} three different versions of \texttt{DCFastQC}, namely (1) \texttt{Hybrid-SE}: \texttt{DCFastQC} with the Hybrid-SE branching (if applicable) and Sym-SE branching {\cheng (otherwise)}, (2) \texttt{Sym-SE}: \texttt{DCFastQC} with the Sym-SE branching only and (3) \texttt{SE}: \texttt{DCFastQC} with the SE branching only. The {\cheng results are} shown in Figure~\ref{fig:comparsion_branching}(a) and (b) for varying $\gamma$ and (c) and (d) for varying $\theta$.  First, both \texttt{Hybrid-SE} and \texttt{Sym-SE} outperform \texttt{SE} {\cheng with} up to 100$\times$ speedup. Moreover, the achieved speedup decreases as $\theta$ {\cheng (resp. $\gamma$)} grows since the search space (i.e., the number of QCs with the size at least $\theta$) narrows with the increase of $\theta$ {\cheng (resp. $\gamma$)}. Second, \texttt{Hybrid-SE} performs the best and achieves around 1 - 5$\times$ speedup compared with \texttt{Sym-SE}. This is well aligned with the theoretical results and demonstrates the efficiency of the Hybrid-SE branching.

\begin{figure}[]
	\centering
	\begin{tabular}{c c}
		\begin{minipage}{3.80cm}
			\includegraphics[width=4.1cm]{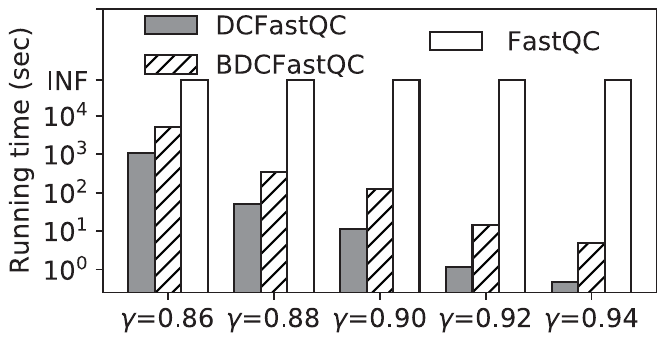}
		\end{minipage}
		&
		\begin{minipage}{3.80cm}
			\includegraphics[width=4.1cm]{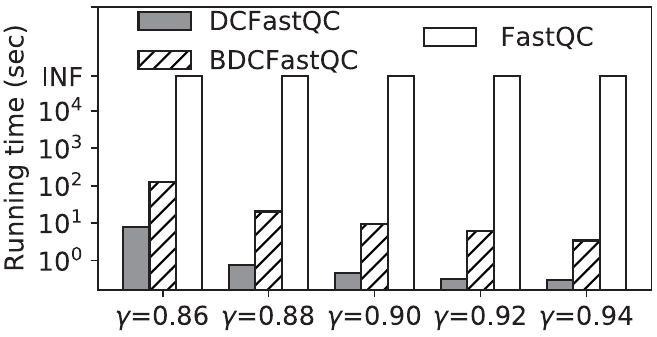}
		\end{minipage}		
		\\
		(a) Varying $\gamma$ (Enron)
		&
		(b) Varying $\gamma$ (Hyves)
		\\
		\begin{minipage}{3.80cm}
			\includegraphics[width=4.1cm]{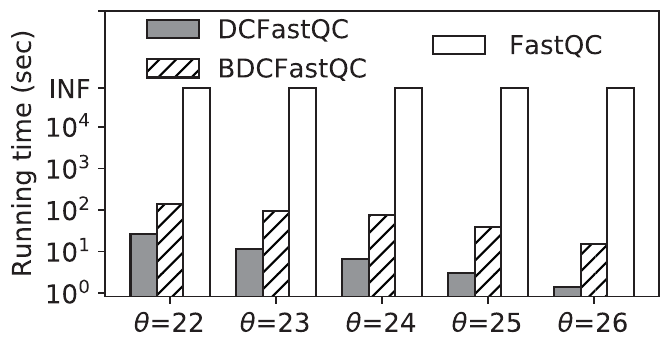}
		\end{minipage}
		&
		\begin{minipage}{3.80cm}
			\includegraphics[width=4.1cm]{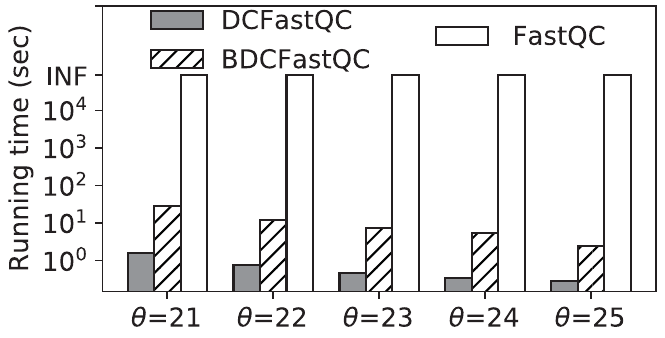}
		\end{minipage}		
		\\
		(c) Varying $\theta$ (Enron)
		&
		(d) Varying $\theta$ (Hyves) 
	\end{tabular}
	\vspace{-0.10in}
	\caption{Comparison among \texttt{DC} frameworks}
	\label{fig:comparsion_framework}
\end{figure}

\smallskip
\noindent\textbf{Comparison among various \texttt{DC} frameworks.} We study the {\cheng effects} of \texttt{DC} frameworks 
{\cheng by comparing}
three different versions, namely, (1) \texttt{FastQC}: without any divide-and-conquer framework, (2) \texttt{BDCFastQC}: with a \emph{basic} divide-and-conquer framework proposed in \cite{khalil2022parallel,guo2020scalable}, (3) \texttt{DCFastQC}: with the \texttt{DC} framework proposed in Section~\ref{sec:DC_Framework}. {\cheng The results are shown in Figure~\ref{fig:comparsion_framework}(a) and (b) for varying $\gamma$ and (c) and (d) for varying $\theta$.} First, \texttt{DCFastQC} and \texttt{BDCFastQC} run significantly faster than \texttt{FastQC} and the achieved speedup increases as $\theta$ or $\gamma$ grows. 
This is well {\cheng aligned} with the theoretical results, i.e., the worst-case running time of \texttt{FastQC} is exponential wrt $n$. Second, \texttt{DCFastQC} outperforms \texttt{BDCFastQC} by achieving at least 10$\times$ speedup. This is because our \texttt{DC} framework with the additional two-hop pruning would produce smaller refined graphs $G_i$ compared with {\cheng those} in~\cite{khalil2022parallel,guo2020scalable}.  

{
\smallskip
\noindent\textbf{Other experiments.} We conduct some additional experiments and put the details in the 
\ifx \CR\undefined
appendix.
\else
technical report~\cite{TR}.
\fi
%
{\cheng (1) We show that the methods that replace the SE branching with our proposed branching methods perform similarly as \texttt{Quick+} does and significantly worse than \texttt{DCFastQC}, which implies that our proposed pruning techniques suit our proposed branching methods better than those in \texttt{Quick+};}
(2) We study the effect of \texttt{DC} on reducing graph size and find that the reduced graph $G_i$ produced by \texttt{DC} is around 0.01\% of the original graph; 
(3) We study the effect of MAX\_ROUND on \texttt{DC} and find that when MAX\_ROUND$=2,3,4$, they would achieve similar performance but better than when MAX\_ROUND$=1$. {\cheng We therefore adopt MAX\_ROUND $=2$ by default.}
}

\section{Related Work}
\label{sec:related}

\noindent\textbf{Maximal quasi-clique enumeration.} In the literature, existing {\cheng studies}~\cite{pei2005mining,jiang2009mining,zeng2006coherent,liu2008effective,khalil2022parallel,guo2020scalable} {\cheng all} adopt a branch-and-bound ({BB}) framework for enumerating MQCs. 
{\cheng They mainly aim to design} effective pruning rules to refine the search space. 
Specifically, \texttt{Crochet}~\cite{pei2005mining,jiang2009mining} and \texttt{Cocain}~\cite{zeng2006coherent} are the earliest {BB} algorithms proposed for mining MQCs. 
They are then combined as a new algorithm \texttt{Quick}~\cite{liu2008effective} which integrates all previous pruning rules and employs new effective ones. 
Authors in~\cite{khalil2022parallel,guo2020scalable} further improve some pruning rules in \texttt{Quick} and {\cheng address} a few boundary cases {\cheng that were} not properly handled before, which leads to the state-of-the-art 
algorithm \texttt{Quick+}. 
To scale \texttt{Quick+} to big graphs, a distributed solution~\cite{guo2020scalable} {\cheng on top of} G-thinker~\cite{yan2022g} and a (single-machine) parallel solution~\cite{khalil2022parallel} {\cheng on top of} T-thinker~\cite{yan2019t} are developed. 
We note that (1) all these {BB} algorithms employ {\cheng the SE branching method} and thus (2) they all have the worst-case time complexity of $O^*(2^{|V|})$. In this paper, we 
{\cheng develop a new {BB} algorithm \texttt{DCFastQC}, which employs new pruning techniques and branching methods and achieves a better time complexity.}

\smallskip
\noindent\textbf{Other variants of quasi-clique mining.} There are many variants of QC mining which consider various problem settings~\cite{chou2015finding,conde2018efficient,lee2016query,sanei2018enumerating,sanei2021mining}, different types of graphs~\cite{ignatov2019preliminary,liu2008quasi,yang2016diversified,lin2021mining,guo2022maximal}, {\cheng and different definitions of QC}~\cite{pattillo2013maximum,conde2018efficient,abello2002massive}. In the sequel, we review these studies.
\underline{First}, 
some 
{\CHENG studies aim}
to only find those QCs that contain a particular vertex~\cite{chou2015finding,conde2018efficient} or a set of query vertices~\cite{lee2016query}. They also adopt a {BB} framework while developing some pruning rules based on the query set.
%
Some other studies aim to find the (top-k) largest QC $G[H]$ such that $|H|$ is maximized~\cite{sanei2018enumerating,sanei2021mining}. In particular, they use a kernel-expansion-based framework. Specifically, to find top-k $\gamma$-QCs, they first find some $\gamma'$-QCs ($\gamma'>\gamma$) as ``kernels'' by using \texttt{Quick}, which are faster to find since $\gamma'>\gamma$. The top-k $\gamma$-QCs are then generated by expanding these kernels. This approach has been shown more efficient than directly mining from the input graph. We note that (1) it still needs to find some $\gamma'$-QCs in the first step by using \texttt{Quick} and (2) it only finds top-k $\gamma$-QCs that contain the kernels.
Therefore, it is hard to adapt these algorithms to improve existing methods for finding all MQCs.
\underline{Second}, 
QC has also {\CHENG been} introduced to bipartite graphs~\cite{ignatov2019preliminary,liu2008quasi}, temporal graphs~\cite{yang2016diversified,lin2021mining} and directed graphs~\cite{guo2022maximal}. Specifically, authors in~\cite{ignatov2019preliminary,liu2008quasi} define quasi-biclique which is a counterpart of QC in bipartite graphs. Besides, temporal quasi-clique is defined on temporal graphs by considering the time
interval that {\cheng a} QC spans over~\cite{yang2016diversified,lin2021mining}. Authors in~\cite{guo2022maximal} introduce directed quasi-clique to directed graphs by considering both the in-degree and out-degree of each vertex. We note that most of {\CHENG these} algorithms are adapted from \texttt{Quick} or \texttt{Quick+} and incorporate additional pruning rules based on specific graph types. 
{\CHENG Hence, these algorithms do not work better than \texttt{Quick+} on general graphs, which are targeted in this paper.}
\underline{Third}, 
{\cheng authors of \cite{pattillo2013maximum,conde2018efficient,abello2002massive} study edge-based QCs, which are different from the degree-based QCs studied in this paper}. Specifically, given a fraction $0\leq \gamma \leq 1$, an edge-based $\gamma$-QC is a subgraph $G[H]$ with the number of edges inside at least $\gamma\cdot |V|(|V|-1)/2$. It has been shown that degree-based QC is denser than edge-based QC~\cite{conde2018efficient}. Therefore, we focus on degree-based QC in this paper. 
Moreover, those algorithms for mining edge-based QCs cannot be adapted to find degree-based QCs 
{\CHENG since these two types of QCs are different.}
In addition, there are some other cohesive subgraphs which tolerate some disconnections inside, {\cheng which include} $k$-plex~\cite{wang2022listing,zhou2020enumerating,dai2022scaling},
$k$-biplex~\cite{yu2021efficient,YuLL022,yu2022maximum},
and $s$-defective clique~\cite{yu2006predicting,chen2021computing,gao2022exact}. However, they {\cheng all satisfy the} hereditary {\cheng property} {\cheng while QCs do not}, and thus their corresponding solutions cannot be adapted to our problem of finding MQCs.

\balance
\section{Conclusion}
\label{sec:conclusion}

%
{\chengrr In this paper,} we propose a new branch-and-bound algorithm \texttt{FastQC} {\chengrr for finding a set of QCs that includes all maximal QCs}. \texttt{FastQC} is based on our developed pruning {\cheng techniques} and branching methods and achieves {\cheng a} smaller worst-case time complexity than the state-of-the-art \texttt{Quick+}. We further develop a divide-and-conquer strategy to boost the performance {\cheng of \texttt{FastQC}}. 
Extensive experiments on real and synthetic datasets {\cheng validate} the superiority of our method. 
In the future, we will develop efficient parallel implementations {\cheng of our algorithms} and 
{\cheng explore possibilities of extending our algorithm to other cohesive subgraph mining problems.}

\begin{acks}
This research is supported by the Ministry of Education, Singapore, under its Academic Research Fund (Tier 2 Award MOE-T2EP20221-0013 and Tier 1 Award (RG77/21)). Any opinions, findings and conclusions or recommendations expressed in this material are those of the author(s) and do not reflect the views of the Ministry of Education, Singapore. The authors would like to thank the anonymous reviewers for providing constructive feedback and valuable suggestions.
\end{acks}
 \balance
\clearpage
\bibliographystyle{ACM-Reference-Format}
\bibliography{SIGMOD_FastQC}

\ifx \CR\undefined
\clearpage
\input{appendix}
\fi

\end{document}